\documentstyle[12pt]{article}

\global\arraycolsep=1pt
\begin{document}

\hfill    SISSA/ISAS 90/94/EP

\hfill    hepth@xxx/9407023

\hfill    July, 1994

\begin{center}
\vspace{10pt}
{\large \bf
MORE ON THE SUBTRACTION ALGORITHM
\, \footnotemark
\footnotetext{Partially supported by EEC,
Science Project SC1$^*$-CT92-0789.}
}
\vspace{10pt}

{\sl Damiano Anselmi}

\vspace{4pt}

International School for Advanced Studies (ISAS), via Beirut 2-4,
I-34100 Trieste, Italia\\
and Istituto Nazionale di Fisica Nucleare (INFN) - Sezione di Trieste,
Trieste, Italia\\

\vspace{8pt}

{\bf Abstract}
\end{center}

\vspace{4pt}

\noindent

We go on in the program of investigating the removal of
divergences of a generical quantum gauge field theory, in the
context of the Batalin-Vilkovisky formalism.
We extend to open gauge-algebr\ae\
a recently formulated algorithm,
based on redefinitions $\delta\lambda$ of the parameters $\lambda$
of the classical Lagrangian and canonical transformations.
The key point is to generalize a well-known
conjecture on the form of the divergent terms to the case of open
gauge-algebr\ae.
We also show that it is possible to
reach a complete control on the effects of the
subtraction algorithm on the space ${\cal M}_{gf}$
of the gauge-fixing parameters. We develop a
differential calculus on ${\cal M}_{gf}$
providing an intuitive
geometrical description of the fact
the on shell physical amplitudes cannot depend on
${\cal M}_{gf}$.
A principal fiber bundle ${\cal E}\rightarrow {\cal M}_{gf}$ with a connection
$\omega_1$ is defined, such that the canonical transformations
are gauge transformations for $\omega_1$.
A geometrical description of the effect of the subtraction
algorithm on the space ${\cal M}_{ph}$ of
the physical parameters $\lambda$ is also
proposed.
At the end, the full subtraction algorithm can be described as a series of
diffeomorphisms on ${\cal M}_{ph}$, orthogonal to ${\cal M}_{gf}$
(under which the action transforms as a scalar), and gauge transformations
on ${\cal E}$. In this geometrical context,
a suitable concept of predictivity is formulated.
Finally, we give some examples
of (unphysical)
toy models that satisfy this requirement, though being neither
power counting renormalizable, nor finite.

\vfill
\eject

\section{Introduction}
\label{intro}
\setcounter{equation}{0}

The Batalin-Vilkovisky formalism \cite{batalin}
provides a useful set-up for the
quantization of Lagrangian field theories. It is now attracting more and
more attention, since the proofs of known results can be considerably
simplified and generalizations are straightforward (see for
example \cite{henneaux,diaz}).

In a recent paper \cite{me}, we formulated, in
the framework of the Batalin-Vilkovisky formalism, an algorithm
for removing the divergences of a quantum gauge field theory.
It preserves a suitable generalization of gauge-invariance
and BRS-invariance, keeps the on-shell physical amplitudes
independent of the gauge-fixing and permits to get a compete
control on the involved arbitrariness. The algorithm is
based on a series of step-by-step
redefinitions of the parameters of the classical Lagrangian
and canonical transformations of fields and BRS sources.

In the present paper, we generalize the results of ref.\ \cite{me}
and give a geometrical description of the subtraction algorithm.
Such a description is expected to be a source of
insight for the classification of predictive quantum field theories.

In ref.\ \cite{me} the starting
theory was supposed to generate a closed gauge-algebra, i.e.\
an off-shell nilpotent BRS operator $s$. Moreover,
the initial BRS action $\Sigma_0$ was assumed to be of the form
\begin{equation}
\Sigma_0(\Phi,K)={\cal L}_{class}(\lambda,\phi)+s\Psi(\Phi)+
K_As\Phi^A.
\label{sigma0}
\end{equation}
In particular, $\Sigma_0$ is $\hbar$-independent and
satisfies both $(\Sigma_0,\Sigma_0)=0$ and $\Delta\Sigma_0=0$
[and, consequently, also the master equation $(\Sigma_0,
\Sigma_0)=2i\hbar \Delta\Sigma_0$].

One of the purposes of this paper is to
show that the results of \cite{me} hold even without
assumption (\ref{sigma0}).

We shall use the same notation that was convenient in ref.\
\cite{me}.  For any further detail, see directly ref.\ \cite{me}.
In (\ref{sigma0}) $\lambda$ denote the parameters
that multiply the gauge-invariant
terms ${\cal G}_i$
of the classical Lagrangian: ${\cal L}_{class}=\sum_i\lambda_i
{\cal G}_i$.
$(\, . \, , \, . \, )$ denote the antibrackets and $\Delta$ is the
Batalin-Vilkovisky delta-operator.
$\phi$ are the classical fields, while $\Phi$ is
the full set of fields, containing
classical fields, ghosts, antighosts, Lagrange multipliers and so on.
$\Psi(\Phi)$ is the gauge-fermion and $K$ are the BRS sources.
The starting action $\Sigma_0$  only depends on $K$ and $\Psi$ {\sl via} the
antifield-combination
\begin{equation}
\Phi_A^*=K_A+{\partial\Psi\over\partial \Phi^A}.
\label{antifield}
\end{equation}
$\Sigma$ denotes the action, while $Z[J,K]$
is the partition function, $W[J,K]$ is
the logarithm of $Z$ and $\Gamma(\Phi,K)$ is the Legendre transform of $W$
with respect to the field sources $J$.
A subscript $k$ marking the functionals
($\Sigma_k$, $Z_k$, $W_k$, $\Gamma_k$ and so on) refers to the
theory in which the divergences have been removed up to the $k^{th}$-loop
order included: $\Gamma_k={\rm finite}+{\cal O}(\hbar^{k+1})$.
Moreover,
$\Gamma^{(k+1)}_{div}$ denotes the $(k+1)^{th}$-loop divergences of
the effective action
$\Gamma_k$:  $\Gamma_k={\rm finite}+\Gamma^{(k+1)}_{div}+
{\cal O}(\hbar^{k+2})$.  $\Gamma^{(k+1)}_{div}$ is a local functional.
$<\ldots>_J$ denotes the average of a functional at nonzero sources $J$.

The operation that removes the order $\hbar^n$ divergences when all lower
order divergences are assumed to have already been removed is denoted by
${\cal L}_n$. The operation ${\cal R}_n={\cal L}_n\circ
{\cal L}_{n-1}\circ\cdots \circ {\cal L}_1$ then removes all the
divergences up to order $\hbar^n$ included.

The property of $\Sigma_0$ of depending on $K$ and $\Psi$ only {\sl via} the
antifield-combination (\ref{antifield}) is not guaranteed to
hold for $\Sigma_k$, $k\geq 1$.
Were it so, the gauge-fixing independence of the physical amplitudes
would be a direct consequence of the master equation \cite{batalin}.
Instead, the master equation only assures the gauge-fixing independence of the
{\sl divergent} amplitudes and the problem is to show that the
subtraction algorithm is compatible with this, so that the finite
physical amplitudes are also gauge-fixing independent.

In \cite{me} it was also useful to define the nilpotent operator
$\sigma={\rm ad }\, \Sigma_0=(\, . \, , \Sigma_0)$, since
$\Gamma^{(n)}_{div}$ was constrained to satisfy
\begin{equation}
\sigma {\Gamma}^{(n)}_{div}=0.
\end{equation}
The general form of $\Gamma^{(n)}_{div}$, solution to the above
equation, was assumed to be
\begin{equation}
\Gamma^{(n)}_{div}={\cal G}^{(n)}(\phi)+\sigma R^{(n)}(\Phi,K),
\label{sigmacoho}
\end{equation}
where ${\cal G}^{(n)}(\phi)=\sum_i\delta_n\lambda_i{\cal G}_i$
is a gauge-invariant functional of
the classical fields $\phi$. Such a conjecture was first proved for
Yang-Mills theories in ref.\
\cite{joglekar}. There, the usual Lorentz gauge was used. However,
the statement is independent on the gauge-choice, since the cohomology
of $\sigma$ is invariant under the canonical transformations that
leave the fields $\Phi$ unchanged, i.e.\
such that the generating functional $F(\Phi,K^\prime)$ has the
form
\begin{equation}
F(\Phi,K^\prime)=K_A^\prime\Phi^A+f(\Phi).
\label{canpec}
\end{equation}
Indeed, due to the fact that the starting action $\Sigma_0$ depends on $K$ and
$\Psi$ only {\sl via} the antifield combination (\ref{antifield}),
the variation
of the gauge-fermion $\Psi(\Phi)\rightarrow \Psi^\prime(\Phi)$
corresponds to a peculiar canonical transformation
of the form (\ref{canpec}) with $f=\Psi-\Psi^\prime$.
The set of canonical transformations of the form (\ref{canpec}) will be called
the {\sl little group} of canonical transformations.

Property
(\ref{sigmacoho}) was assumed by Stelle in \cite{stelle} for higher
derivative quantum gravity. It has been recently proved for matter coupled
Yang-Mills theory with a semi-simple gauge-group in ref.\ \cite{henneaux},
where some peculiar features of abelian gauge groups are also pointed out.

Notice that
conjecture (\ref{sigmacoho}) is not independent of the choice of variables,
since it is not invariant under the most general
canonical transformation. When quantizing a classical theory,
a ``boundary condition'' \cite{batalin}
is imposed on the action $\Sigma$ when solving the master
equation
\begin{equation}
(\Sigma,\Sigma)=2i\hbar \Delta\Sigma,
\label{masterequation}
\end{equation}
namely the condition that the order zero part
$S$ of the action $\Sigma$ (let us write
$\Sigma=S+\sum_{n=1}^\infty \hbar^n M_n$)
reduces to the classical Lagrangian
${\cal L}_{class}(\phi,\lambda)$,
when the antifields $\Phi^*$ are set to zero:
\begin{equation}
S|_{\Phi^*=0}={\cal L}_{class}(\phi,\lambda).
\label{boundary}
\end{equation}
This condition is invariant
under the little group of canonical transformations,
but not under the full group, precisely as (\ref{sigmacoho}).
Thus, condition (\ref{sigmacoho}) is a statement about the
compatibility of the variables that appear in
${\cal G}^{(n)}$ and in the boundary condition (\ref{boundary}).

In section \ref{geometry} we shall give a geometrical
interpretation of (\ref{sigmacoho}) according to which
${\cal G}^{(n)}(\phi)$ is the Lie derivative of $S$ along
a vector field tangent to the space of the parameters $\lambda$
and orthogonal to the space of gauge-fixing parameters.
Compatibility of (\ref{sigmacoho}) and (\ref{boundary})
then means that the content
of the $\sigma$-cohomology should be fully encoded in the classical
Lagrangian ${\cal L}_{class}$.

The meaning of (\ref{sigmacoho})
is that  the divergent terms can be of two types:
the ``Lagrangian type'' terms
${\cal G}^{(n)}(\phi)$,
that are removed by redefining the parameters $\lambda$ of
the classical Lagrangian and the ``gauge-fermion type'' terms
$\sigma R^{(n)}(\Phi,K)$,
that are removed by a canonical transformation on fields $\Phi$
and BRS sources $K$, with a generating functional
\begin{equation}
F^{(n)}(\Phi,K^\prime)=K_A^\prime\Phi^A+R^{(n)}(\Phi,K^\prime).
\label{fn}
\end{equation}
In ref.\ \cite{me} we proved a ``no-mixing'' theorem,
that states that the redefinitions
of $\lambda$ are independent of the
gauge-fermion $\Psi$.
In other words, the gauge-fermion does not mix with
the classical Lagrangian,
although the classical Lagrangian mixes with the gauge-fermion
(i.e.\ $R^{(n)}$ depends on $\lambda$, although
${\cal G}^{(n)}$ does not depend on $\Psi$).
This fact assures that the on-shell
physical amplitudes are independent of the gauge-fixing parameters.

We identified the following sequence of generalizations of
gauge-transformations
\begin{equation}
\delta_{gauge}\rightarrow s\rightarrow \sigma
\rightarrow \Omega \rightarrow {\rm ad}\, \Gamma.
\label{sequence}
\end{equation}
Every operator, except for the gauge-transformation operator
$\delta_{gauge}$, is nilpotent:
\begin{equation}
\matrix{s^2=0,& \sigma^2=0,&\Omega^2=0,&({\rm ad}\, \Gamma)^2=0}.
\end{equation}
The BRS operator
$s$ only acts on the fields $\Phi$ and not on the BRS sources $K$.
For closed gauge-algebr\ae\ it
coincides with $\delta_{gauge}$ on the classical fields $\phi$.
$\sigma$ is a generalization of $s$ to the space of fields and BRS sources.
It coincides with $s$ on the fields $\Phi$. As we have
anticipated, the cohomological content of the operator $\sigma$
is invariant under the little group of
canonical transformations (\ref{canpec}).
$\Omega={\rm ad}\, \Sigma-i\hbar \Delta$  is a further generalization
of $\sigma$, in the sense that the $\Omega$-cohomology is
invariant under the most general canonical transformation.
All the previous operators act on ``integrand functionals".
The operator ${\rm ad}\, \Gamma$ represents
a generalization of $\Omega$ acting on ``integrated functionals", i.e.\
the average values of the ``integrand functionals".

The nice feature of the algorithm is that
what happens is clearly {\sl visible},
even when some nonrenormalizable
vertices\footnotemark
\footnotetext{To avoid misunderstandings, our concept of
nonrenormalizability refers to power-counting
nonrenormalizability \cite{itzykson}.} are present
(of whatever origin: exotic gauge-fixing, highly dimensioned
composite operators, genuine nonrenormalizability of the theory).
We showed that the useful formul\ae\ holding
for the initial action $\Sigma_0$
can be naturally ``propagated'' ({\sl via} the
subtraction algorithm) to the ``renormalized'' action $\Sigma_\infty$
and to the $\Gamma$-functional.

We noticed that the algorithm can be viewed as a way
of implementing the
{\sl principle of correspondence}: one starts from some convenient
{\sl classical} variables and
looks for the
{\sl correct} quantum variables and quantum parameters.
The assumptions made in ref.\ \cite{me} assure that
the divergences can be reabsorbed by simply redefining
those quantities (fields, parameters and sources)
that are already present at the classical level
\footnotemark\footnotetext{
The whole initial BRS action $\Sigma_0$
can be considered {\sl classical}, since
it is $\hbar$-independent.}: the idea is that
nothing can be added, but anything
can be adjusted.

The requirement of a precise {\sl correspondence}
between classical and quantum
worlds is quite natural: indeed, since we are only able to explore the
quantum world by means of classical instruments,
what we observe is only the part of the quantum world
that has a correspondence with the classical one. Since
we are not allowed to say that there exists more than what we see,
we conclude that the whole quantum
world is in correspondence with the classical one.

The first purpose of this paper is to show that some of the assumptions
that were made in ref.\ \cite{me} can be relaxed, i.e.\ that the
subtraction algorithm can be applied in more general cases than those
considered in ref.\ \cite{me}. For example, the results also apply to
open gauge-algebr\ae, i.e.\ gauge
field theories such that the BRS operator $s$ is not off-shell nilpotent,
rather
$s^2=0$ only on shell. In this case, the starting action $\Sigma_0$
is not linear in $K$, so that it is not possible to restrict to theories in
which $\Sigma_0$ has the form (\ref{sigma0}). Moreover,
$\Sigma_0$ is not required to satisfy both
$(\Sigma_0,\Sigma_0)=0$ and $\Delta\Sigma_0=0$ separately, but
only the full master equation $(\Sigma_0,\Sigma_0)=
2i\hbar \Sigma_0$. The order $\hbar^0$ part of this equation
gives
\begin{equation}
(S,S)=0,
\label{clmast}
\end{equation}
i.e.\ the order $\hbar^0$ part $S$ of the starting action $\Sigma_0$
satisfies the so-called ``classical master equation". The correct definition
of the operator $\sigma$ is then
\begin{equation}
\sigma={\rm ad}\, S,
\label{sigma}
\end{equation}
instead of ${\rm ad}\, \Sigma_0$. Indeed $\sigma^2=0$ is assured
by (\ref{clmast}) and not
by the master equation (which assures the nilpotency of $\Omega$).
Moreover,
the BRS operator $s$ corresponds to
\begin{equation}
s\Phi^A=\left.{\partial_l S\over \partial K_A}\right|_{K=0}.
\label{brs}
\end{equation}
We see that $\sigma$ no more coincides
with the BRS operator $s$ on the fields $\Phi$, rather,
$s\Phi^A=\sigma \Phi^A |_{K=0}$. Consequently,
the classical Lagrangian ${\cal L}_{class}=\sum_i\lambda_i{\cal G}_i$
is in general not $\sigma$-closed. This fact implies that a suitable
generalization of conjecture (\ref{sigmacoho}) should be found.
All this is extensively discussed in section \ref{generalization},
where we show explicitly that any
argument of ref.\
\cite{me} can be adapted to the case of open algebr\ae\
[or to the case of closed algebr\ae\ such that $\Sigma_0$ has
not the form (\ref{sigma0})].
Provided the above remarks are taken into account,
sequence (\ref{sequence}) and its properties are unaltered.

The second purpose of this paper is to search for a geometrical description
of the subtraction algorithm, since it could be a source of insight
in the program of classifying those
nonrenormalizable quantum field theories that are predictive.
We develop arguments stressing the fact that our
subtraction algorithm extends very
easily suitable identities that hold for the initial action $\Sigma_0$ to
any step of the subtraction procedure and so to the convergent
effective action $\Gamma_\infty$.
In other words, the algorithm is {\sl tractable}
even when (like in the case of nonrenormalizable theories)
the structure of the counterterms is very complicated:
it is possible to isolate the meaningful properties of the subtraction
procedure from the contingent complications.

We develop a complete differential calculus on the manifold ${\cal M}_{gf}$
of the gauge-fixing parameters (section \ref{covariant}), permitting
to reach a complete control of the effects of the subtraction algorithm on the
gauge-fixing sector. This is achieved by introducing
a certain set of differential
forms $\omega_k$ of degrees $k=0,\ldots m$ on ${\cal M}_{gf}$,
the zero-form $\omega_0$ being the effective action
$\Gamma$.
$\omega_k$ satisfy a certain set of differential
identities ({\sl cascade equations}) that start with the Ward
identity $(\Gamma,\Gamma)=0$ and that are
preserved by the subtraction algorithm. In section \ref{application}
we show that
the cascade equations imply descent equations  for the divergent
parts $\omega_{k\, div}$ of $\omega_k$ and
this property permits to choose
the canonical transformation (\ref{fn})
so as to make any $\omega_k$
convergent, while preserving the gauge-fixing independence of
the redefinitions $\lambda_i-\delta_n\lambda_i$
of the parameters $\lambda_i$
(``no mixing theorem'').

In ref.\ \cite{me} this was shown focusing on
{\sl one} gauge-fixing parameter $\kappa$ only, instead of a
generic manifold ${\cal M}_{gf}$.
Precisely, a functional $S_n=<\chi_n>_J$ such that
\begin{equation}
\matrix{
{\partial \Gamma_n\over \partial \kappa}=(S_n,\Gamma_n),\quad &\quad
{\partial\Sigma_n\over \partial \kappa}=\Omega_n\chi_n,}
\label{usefulid}
\end{equation}
was introduced.
Initially, the gauge-fixing parameter $\kappa$ only enters in
the gauge-fermion: $\Psi=\Psi(\Phi,\kappa)$. $\chi_n$ is
a local functional and its order zero part coincides
with ${\partial\Psi\over \partial \kappa}$.
We showed that it is possible to choose the canonical transformations
(\ref{fn}) so as to make $S_n$ convergent to order $\hbar^n$
(provided $S_{n-1}$ is inductively assumed to be
convergent up to $\hbar^{n-1}$).
Due to this, equations (\ref{usefulid})
can be extended to any step of the subtraction procedure and permit
to show that the on-shell physical amplitudes are $\kappa$-independent.

We gave two different proofs of the above
property, stressing different
features of the subtraction algorithm. The first method has the
property that it can be easily generalized to a generic manifold
${\cal M}_{gf}$, but its disadvantage
is that it is only applyable in the context of a regularization technique,
like the dimensional one, where it is possible to set $\Delta=0$
on local functionals. The second method, instead, does not
suffer from this restriction, but its generalization to a manifold
${\cal M}_{gf}$ is less straightforward. This fact can be turned
into a positive feature, since it leads to the development of
the interesting differential calculus on ${\cal M}_{gf}$
that we anticipated.

In section \ref{geometry}, we give
a geometrical description of the subtraction
algorithm and of the independence
from the gauge-fixing parameters.
We define a principal fiber bundle ${\cal P}$, whose
fiber is isomorphic to the group of canonical transformations and whose
base manifold is ${\cal M}_{gf}$. The form $\omega_1$ is a connection
on ${\cal P}$ and its field strength vanishes on shell. The canonical
transformations are the gauge transformations for $\omega_1$.

We also give the building blocks for the geometrical
description of the effects
of the subtraction algorithm on the space ${\cal M}_{ph}$ of
the physical parameters $\lambda$.
The set of redefinitions $\delta \lambda$ is viewed as a diffeomorphism
on ${\cal M}_{ph}$, orthogonal to ${\cal M}_{gf}$.
We translate
into this language the requirement that finitely many parameters
are sufficient for removing all the
divergences, by formulating a suitable concept of predictivity.
Finally, in section \ref{toy}
we investigate how this could happen, by means of suitable toy
Lagrangians that are not power counting renormalizable.

Let us write $S_n^{(\kappa)}$, $\chi_n^{(\kappa)}$ and so on,
when it is convenient to denote
the gauge-fixing parameter explicitly.
We introduce a useful concept of ``covariant derivative"
$D_\kappa$ with respect to
the gauge-fixing parameter $\kappa$,
\begin{equation}
D_\kappa={\partial\over \partial \kappa}-{\rm ad}_l\,
{\partial \Psi\over \partial \kappa},
\label{covder}
\end{equation}
where ${\rm ad}_l$ denotes the left-adjoint operation, i.e.\
${\rm ad}_l\, X\, Y=(X,Y).$
The nice property of $D_\kappa$ is that it sends $\sigma$-closed
and $\sigma$-exact functionals into $\sigma$-closed
and $\sigma$-exact functionals, i.e.\ it commutes with $\sigma$:
\begin{equation}
[\sigma,D_\kappa]=0.
\label{commut}
\end{equation}
This fact is a simple consequence of the covariant constancy of the
order $\hbar^0$-part $S(\Phi,K)$ of
the starting action $\Sigma_0(\Phi,K)$:
\begin{equation}
D_\kappa S=0.
\label{covar}
\end{equation}
It is also simple to prove that the manifold ${\cal M}_{gf}$
of the gauge-fixing
parameters is flat with respect to the above covariant derivative,
namely
\begin{equation}
[D_\kappa,D_\alpha]=0,
\label{dquadro}
\end{equation}
for any couple of gauge-fixing parameters $\kappa$ and $\alpha$.

Let $\{\kappa_1,\ldots,\kappa_m\}$
denote the set of gauge-fixing parameters which the gauge-fermion $\Psi$
depends on ($m={\rm dim}\, {\cal M}_{gf}$).
We define an exterior derivative on ${\cal M}_{gf}$:
\begin{equation}
d=d\kappa_i \left.{\partial\over\partial\kappa_i}\right|_{\Phi,K}.
\label{dgf}
\end{equation}
Then, the covariant derivative (\ref{covder}) and properties
(\ref{commut}), (\ref{covar}) and (\ref{dquadro}) read
\begin{equation}
D=d-{\rm ad}_l\, d\Psi,\quad\quad
[D,\sigma]=0,\quad\quad  D^2=0,\quad\quad DS=0.
\label{Dcov}
\end{equation}
We can also define the connection
\begin{equation}
\omega_n=S_n^{(\kappa_i)}dk_i
\label{opi}
\end{equation}
and the curly covariant derivative
\begin{equation}
{\cal D}_n=d-{\rm ad}_l\,\omega_n.
\end{equation}
{}From (\ref{usefulid}) we have
\begin{equation}
{\cal D}_n\Gamma_n=0,
\end{equation}
that generalizes $DS=0$. We see that $\Gamma$ is covariantly constant
with respect to the curly covariant derivative.

\section{Extension to open gauge-algebr\ae}
\label{generalization}
\setcounter{equation}{0}

In this section we show that the results of ref.\ \cite{me}
can be extended to theories corresponding to open gauge-algebr\ae\
as well as
to closed gauge-algebr\ae\ such
that  $\Sigma_0$ is not of the form (\ref{sigma0}).
It is useful to stress the assumptions that are
relaxed and those that are maintained.
The gauge-fermion $\Psi$ is still supposed to be a functional of
the fields $\Phi$ only and not of the BRS sources $K$. It may depend on some
gauge-fixing parameters $\kappa_i$, but we assume that it
is independent of the parameters $\lambda$ of the
classical Lagrangian. Any such dependence would represent an
undesired identification between gauge-fixing parameters and Lagrangian
parameters: eventually, it is better to make this identification
at the very end, i.e.\ in the final convergent theory.
In this way, one has a clearer perception of what
happens in the subtraction procedure.
Moreover, $\Psi(\Phi,\kappa_i)$ is assumed to be local, convergent,
independent of $\hbar$ and such as to produce standard propagators
(i.e.\ gauge-conditions, like the axial-gauge, that produce
non-local divergent counterterms, should be treated apart).
$\Psi$ is not constrained by power-counting requirements.

The starting action $\Sigma_0(\Phi,K)$ is assumed
to satisfy the master equation
\begin{equation}
(\Sigma_0,\Sigma_0)=2i\hbar\Delta\Sigma_0.
\end{equation}
Moreover, $\Sigma_0$ depends on the BRS sources $K$ and the
gauge-fermion $\Psi$ only {\sl via} the antifield-combination
(\ref{antifield}),
so that the derivative of $\Sigma_0$ with respect to the gauge-fixing
parameters is
\begin{equation}
d\Sigma_0=
d\left({\partial\Psi\over \partial\Phi^A}\right)
{\partial_l\Sigma_0\over \partial K_A}=
(d\Psi,\Sigma_0)=\Omega_0 d\Psi.
\label{desigma0sudek}
\end{equation}
One can also write $D\Sigma_0=0$, with $D$ being defined by (\ref{Dcov}).
In general, let $\Sigma_0=S+\sum_{n=1}^\infty\hbar^n M_n$ be the
expansion of $\Sigma_0$ as a power series in $\hbar$.
$S$ satisfies the classical master equation (\ref{clmast}) and the boundary
condition (\ref{boundary}). The classical Lagrangian ${\cal L}_{class}=
S|_{\Phi^*=0}$ is written as
\begin{equation}
{\cal L}_{class}=\sum_i\lambda_i{\cal G}_i,
\label{l1}
\end{equation}
where ${\cal G}_i$ is a basis of gauge-invariant functionals of the
classical fields: $\delta_{gauge}{\cal G}_i=0$. The BRS
operator $s$ is defined by (\ref{brs})
and $\sigma$ by (\ref{sigma}).
This means that it is no longer true that $\sigma {\cal G}_i=0$:
${\cal G}_i$ cannot represent a basis of the $\sigma$ cohomology
and conjecture (\ref{sigmacoho}) has to be suitably
generalized. In a moment we shall see how it is reasonable to extend it.

Since $\Psi$ is assumed to be $\hbar$-independent, $S$ also
depends on $K$ and $\Psi$ {\sl via} the antifield-combination
(\ref{antifield}).
It satisfies the classical master equation (\ref{clmast}) and
the property $DS=0$
(\ref{covar}), that can be derived taking the order $\hbar^0$
part of (\ref{desigma0sudek}).
As anticipated, we no longer introduce assumptions on the form of $\Sigma_0$
or $S$ or on the BRS invariance of the functional measure.

In the remainder of the section we generalize the arguments of ref.\
\cite{me}. However, we do not repeat
the complete derivations, for brevity.
First notice that, under the inductive assumption
that $\Gamma_{n-1}$ is finite
up to order $\hbar^{n-1}$ included, the Ward identity
\begin{equation}
(\Gamma_{n-1},\Gamma_{n-1})=0,
\end{equation}
assures that
\begin{equation}
(\Gamma^{(n)}_{div},S)=\sigma \Gamma^{(n)}_{div}=0.
\label{divequation}
\end{equation}
As anticipated, conjecture (\ref{sigmacoho}) is no more correct,
since ${\cal G}_i$
are not $\sigma$-closed, but only $s$-closed.
Differentiating (\ref{clmast}) with respect to $\lambda_i$, we obtain
\begin{equation}
({\cal S}_i,S)=\sigma {\cal S}_i=0,
\end{equation}
where ${\cal S}_i\equiv {\partial S\over \partial \lambda_i}$.
Consequently, it is natural to conjecture that ${\cal S}_i$ is a basis
of the $\sigma$-cohomology, i.e.\ that
the most general solution
to equation (\ref{divequation}) is
\begin{equation}
\Gamma^{(n)}_{div}=\sum_i \delta_n\lambda_i {\cal S}_i+
\sigma R^{(n)}.
\label{newconj}
\end{equation}
We shall still denote the sum
$\sum_i \delta_n\lambda_i {\cal S}_i$ by ${\cal G}^{(n)}$. Notice that
\begin{equation}
{\cal S}_i|_{\Phi^*=0}={\cal G}_i
\end{equation}
and that ${\cal S}_i$ do not necessarily depend on the classical fields only,
but can also depend on the other fields and the BRS sources. Clearly,
they depend on the BRS sources $K$ and the gauge-fermion $\Psi$
only {\sl via} the combination (\ref{antifield}).
Moreover, although $d {\cal G}_i=0$,
since no gauge-fixing is involved in the determination of the
basis of gauge-invariant functionals
${\cal G}_i$, the same is no longer true for ${\cal S}_i$, rather,
as in formula (\ref{desigma0sudek}),
\begin{equation}
 D  {\cal S}_i=d{\cal S}_i
-(d\Psi,{\cal S}_i)=0,
\label{desisudek}
\end{equation}
a result that can be also obtained by differentiating
$DS=0$ with respect to $\lambda_i$ and noticing that
${\partial d\Psi\over\partial\lambda_i}=0$. So, ${\cal S}_i$
are covariantly constant with respect to $D$.

If (\ref{newconj}) holds, then it is possible to proceed as in the
case of closed gauge-algebr\ae: ${\cal G}^{(n)}$ is removed by a redefinition
$\lambda_i\rightarrow \lambda_i-\delta_n\lambda_i=
\lambda_i+{\cal O}(\hbar^n)$ of the
parameters $\lambda_i$ of the classical Lagrangian. Indeed,
\begin{eqnarray}
\Sigma_{n-1}(\Phi,K,\lambda_i-\delta_n\lambda_i)&=&
\Sigma_{n-1}(\Phi,K,\lambda_i)
-\sum_i\delta_n\lambda_i{\partial\Sigma_{n-1}\over \partial\lambda_i}
+{\cal O}(\hbar^{n+1})\nonumber\\&=&
\Sigma_{n-1}(\Phi,K,\lambda)
-\sum_i\delta_n\lambda_i{\cal S}_i+{\cal O}(\hbar^{n+1}).
\end{eqnarray}
Then, $\sigma R^{(n)}$ is removed by a canonical transformation
generated by (\ref{fn})
with no difference with respect to the case considered in ref.\ \cite{me}.

Let us now present the modifications to the proof of independence from
the gauge-fermion, i.e.\
\begin{equation}
d\delta_m\lambda_i=0
\hskip .3truecm \forall m
\hskip .3truecm \forall i.
\label{op}
\end{equation}
This was done, in ref.\ \cite{me}, for a single gauge-fixing parameter
$\kappa$, not for a generic manifold ${\cal M}_{gf}$. In this section,
we generalize the proof of ref.\ \cite{me} for a single parameter
to the case of
open-algebr\ae. In the next sections, we generalize the proof
to the full manifold ${\cal M}_{gf}$.

Assuming inductively that (\ref{op}) holds up to $m=n-1$,
we can extend
equation (\ref{desigma0sudek}) to
\begin{equation}
{\partial \Sigma_{n-1}\over \partial \kappa}=\Omega_{n-1}
\chi^{(\kappa)}_{n-1},
\label{rev1}
\end{equation}
for some local functional $\chi^{(\kappa)}_{n-1}$, whose zeroth order part
is ${\partial\Psi\over\partial\kappa}$.
Consequently, we also have \cite{me}
\begin{equation}
{\partial\Gamma_{n-1}\over \partial\kappa}=(S^{(\kappa)}_{n-1},\Gamma_{n-1}),
\label{degammasudek}
\end{equation}
where $S^{(\kappa)}_{n-1}=<\chi_{n-1}^{(\kappa)}>_J$.
$S^{(\kappa)}_{n-1}$ is inductively assumed to be finite up
to order $\hbar^{n-1}$:
$S_{n-1}^{(\kappa)}={\rm finite}+S^{(n)}_{\kappa\, div}+
{\cal O}(\hbar^{n+1})$, $S^{(n)}_{\kappa\, div}$ denoting the
order $\hbar^n$-divergent part, which is local.

The order $\hbar^n$ divergent part of equation (\ref{degammasudek}) gives,
as in ref.\ \cite{me},
\begin{equation}
{\partial \Gamma^{(n)}_{div}\over\partial\kappa}=\left(
{\partial \Psi\over \partial \kappa},
\Gamma^{(n)}_{div}\right)+\sigma S_{\kappa\, div}^{(n)},
\end{equation}
that can be also written as
\begin{equation}
 D _\kappa\Gamma^{(n)}_{div}=\sigma S_{\kappa\, div}^{(n)}.
\end{equation}
Using (\ref{newconj}) and (\ref{commut}) we have
\begin{equation}
 D _\kappa {\cal G}^{(n)}=\sigma S_{\kappa\, div}^{(n)}- D _\kappa
\sigma R^{(n)}=\sigma (S_{\kappa\, div}^{(n)}- D _\kappa R^{(n)}).
\end{equation}
At this point, equation (\ref{desisudek}) permits to write
\begin{equation}
 D _\kappa {\cal G}^{(n)}= D _\kappa \left(\sum_i\delta_n\lambda_i
{\cal S}_i\right)=\sum_i{\partial \delta_n\lambda_i
\over\partial \kappa}
{\cal S}_i=\sigma (S_{\kappa\, div}^{(n)}- D _\kappa R^{(n)}).
\label{both}
\end{equation}
Since by assumption ${\cal S}_i$ is a basis of the $\sigma$ cohomology,
we conclude that both sides of equation (\ref{both}) vanish, so that
\begin{equation}
{\partial \delta_n\lambda_i\over\partial \kappa}=0 \hskip .3truecm
\forall i.
\label{indep}
\end{equation}
This assures that (\ref{rev1}) and (\ref{degammasudek})
can be extended to order $\hbar^n$, giving (\ref{usefulid}).

Finally, in order to fully reproduce the inductive assumptions,
one also has to
prove that $S_n^{(\kappa)}$ can be chosen finite up to
order $\hbar^n$ included.
This can be done exactly as in ref.\ \cite{me}
by means of a suitable choice of the functional
$R^{(n)}$
in the canonical transformation $F^{(n)}$ (\ref{fn}).

We recall that
(\ref{newconj}) determines
$R^{(n)}$ only up to additions of $\sigma$-closed functional $T^{(n)}$:
\begin{equation}
R^{(n)}\rightarrow R^{(n)}+T^{(n)}, \hskip .5truecm \sigma T^{(n)}=0.
\label{freedom}
\end{equation}
A corresponding freedom characterizes the canonical transformation
(\ref{fn}).
If $T^{(n)}={\cal O}(\hbar^n)$,
such additions do not change the action $\Sigma_n$ up to order
$\hbar^n$ included, so that the order $\hbar^n$-change
of the average of a functional can only be due to the change of
the functional and not to the change of the average. Clearly,
$\Gamma_n$ is convergent up to order $\hbar^n$ for any $T^{(n)}
={\cal O}(\hbar^n)$.

We noticed in ref.\ \cite{me}
that under the canonical transformation generated by
$F^{(n)}$, $\chi_{n-1}^{(\kappa)}$ changes into
\begin{equation}
\tilde\chi_{n-1}^{(\kappa)}-{\partial F^{(n)}\over\partial \kappa}.
\label{chitr}
\end{equation}
We recall \cite{me} that the tilde means that the old variables
$\{\Phi,K\}$ have to be replaced with the new ones
$\{\Phi^\prime,K^\prime\}$, considered as functions of the old ones:
$\{\Phi^\prime(\Phi,K),K^\prime(\Phi,K)\}$.
$\chi_n^{(\kappa)}$ is thus obtained from $\chi_{n-1}^{(\kappa)}$
by letting $\lambda_i\rightarrow \lambda_i-\delta_n\lambda_i$
and performing the canonical transformation (\ref{fn})
according to (\ref{chitr}).

Let us assume that $\Gamma_{div}^{(n)}$ has been removed
and $T^{(n)}$ is of order $\hbar^n$ and
divergent.
Under a variation (\ref{freedom}) of the generating functional
(\ref{fn}), $\chi_n^{(\kappa)}$ varies of
$- D _\kappa T^{(n)}+{\cal O}(\hbar^{n+1})$, so that
the order $\hbar^n$ divergent part ${\cal S}^{(n)}_{\kappa\, div}$
of $S_n^{(\kappa)}=<\chi_n^{(k)}>_J$ varies as
\begin{equation}
{\cal S}^{(n)}_{\kappa\, div}\rightarrow {\cal S}^{(n)}_{\kappa \,
div}- D _\kappa
T^{(n)}.
\end{equation}
We also recall that ${\cal S}^{(n)}_{\kappa\, div}$ is $\sigma$-closed
\begin{equation}
\sigma {\cal S}^{(n)}_{\kappa\, div}=0,
\label{closure}
\end{equation}
due to the first equation
of (\ref{usefulid}), i.e.\
${\partial\Gamma_n\over\partial\kappa}=(S_n^{(\kappa)},\Gamma)$.
Given ${\cal S}^{(n)}_{\kappa \, div}$, the condition
\begin{equation}
{\cal S}^{(n)}_{\kappa \, div}= D _\kappa T^{(n)}
\label{eqa}
\end{equation}
for $T^{(n)}$  is solved (perturbatively in $\kappa$) by
\begin{equation}
T^{(n)}=\sum_{j=1}^\infty {(-1)^{j+1}\kappa^j\over j!} D_\kappa^{j-1}
{\cal S}^{(n)}_{\kappa\, div}.
\label{eqb}
\end{equation}
Due to (\ref{commut}) and (\ref{closure}), $T^{(n)}$ is $\sigma$-closed,
as desired. Moreover, it is order $\hbar^n$ and divergent.
The above choice of $T^{(n)}$ produces a functional
$S_n^{(\kappa)}$ that is convergent up to order $\hbar^n$ included.

We conclude noticing that if the starting action $\Sigma_0(\Phi,\Phi^*)$
is power counting renormalizable and the most general
power counting renormalizable classical Lagrangian
${\cal L}_{class}(\lambda,\Phi)$ depends
on a finite number of parameters $\lambda$,
then the above result combined
with power counting assures that the quantum theory is
predictive, namely that all the divergences are removed by
canonical transformations and redefinitions of the parameters $\lambda$
and that the on-shell physical amplitudes depend on a finite number
of parameters. In other words, the fact that the algebra is open
has no dramatic consequence, provided the
generalization (\ref{newconj}) of the usual conjecture
(\ref{sigmacoho}) holds.

\section{Covariant treatment of the gauge-fixing parameters}
\label{covariant}
\setcounter{equation}{0}

In this section, we develop a differential calculus on ${\cal M}_{gf}$. We
define differential forms $\omega_k$ of degree
$k=1,\ldots m={\rm dim}\,{\cal M}_{gf}$, the zero-form $\omega_0$
being the effective action $\Gamma$ and the one form $\omega_1$
being (\ref{opi}).
$\omega_k$ satisfy certain improved descent equations, called by
us {\sl cascade equations} due to their \ae sthetic aspect,
that are preserved
by the subtraction algorithm and possess some {\sl cascade invariances}
that will be
fundamental in section \ref{application}, where we shall apply these
properties to show that $F^{(n)}$ (\ref{fn})
can be chosen in order to make any
$\omega_k$ finite up to order $\hbar^n$ and $d\delta_n\lambda_i=0$
$\forall i$. In section \ref{geometry}
we shall define a fiber bundle ${\cal P}$ on ${\cal M}_{gf}$
that provides an intuitive geometrical description of the fact that
the subtraction algorithm is not able to pick up any
information from ${\cal M}_{gf}$. $\omega_1$ is a connection on ${\cal P}$.

Our purpose is to generalize the construction of the previous section
to the
$m$-dimensional manifold ${\cal M}_{gf}$.
In the remainder, apart from the situations of possible
misunderstanding, we omit the suffix $n$ in $\Gamma_n$, $\Sigma_n$,
$\chi_n$, $S_n$, and so on.

It is easy to prove, from
equation (\ref{chitr}), the following formula
\begin{equation}
d\chi={1\over 2}(\chi,\chi).
\label{dechi}
\end{equation}
Indeed, this equation holds for $n=0$ and is canonically preserved
(this fact will be a straightforward consequence of a computation
that will be made later on).
Moreover, due to (\ref{indep}),
the redefinitions of the parameters $\lambda$
do not affect it, so that (\ref{dechi}) is also preserved by ${\cal L}_n$ and
by the operation ${\cal R}_n$
that removes the divergences up to order $n$ included. This proves
(\ref{dechi}).

Let us write $\omega_{-1}=0$, $\omega_0=\Gamma$,
$\omega_1=<\chi>_J$ (now
$\omega_1$ denotes what we previously called $\omega_n$).
We have, from the first of (\ref{usefulid}),
\begin{equation}
d\omega_0=(\omega_1,\omega_0).
\label{uopo}
\end{equation}
We want to prove that
\begin{equation}
d\omega_1={1\over 2}(\omega_1,\omega_1)-(\omega_2,\omega_0),
\label{deomega1}
\end{equation}
where
\begin{equation}
\omega_2=
{i\over 2\hbar}<\chi\chi>_J-{i\over2\hbar}\omega_1\omega_1
-{1\over 2}\{\omega_1,\omega_1\}.
\label{omega2}
\end{equation}
$\omega_2$ is a two-form on ${\cal M}_{gf}$.
The symbol of wedge product among forms on ${\cal M}_{gf}$ is
understood. The curly brackets do not denote an anticommutator
but a different notion of brackets, that is convenient to analyse
explicitly. Precisely,
\begin{equation}
\{X,Y\}=\left. {\partial_r X\over \partial\Phi^A}\right|_K
\left.{\partial_l Y\over \partial J_A}\right|_K.
\label{mixedbrackets}
\end{equation}
They will be called {\sl mixed brackets}, due to the fact that they mix
derivatives with respect to $\Phi$ and $J$, while $K$ is
kept constant. They possess a nice diagrammatical meaning, that will be
illustrated in the sequel.
It is easy to prove that the mixed brackets satisfy the following
properties
\begin{equation}
\matrix{
\{X,Y\}=(-1)^{\varepsilon(X)\varepsilon(Y)+d_X d_Y}\{Y,X\},\cr
\{X,YZ\}=\{X,Y\}Z+(-1)^{\varepsilon(Y)\varepsilon(Z)+d_Y d_Z}\{X,Z\}Y.}
\end{equation}
Here $d_X$ denotes the form degree of $X$. The factors like $(-1)^{d_X d_Y}$
are due to the fact that $X$ and $Y$ have been interchanged.
One must also keep into account
that when applying identities for the antibrackets to differential forms
on ${\cal M}_{gf}$, similar corrections involving the form
degrees are necessary whenever the order of the forms is changed.

Another property of the mixed brackets (\ref{mixedbrackets}) is
\begin{equation}
(X,Y)+\{X,(\omega_0,Y)\}+\{(X,\omega_0),Y\}=
(-1)^{\varepsilon(X)}(\omega_0,\{X,Y\}).
\label{identity}
\end{equation}
The proof of this identity is more involved and will be given explicitly.

Let us write the antibrackets in the form
\begin{eqnarray}
(X,Y)&=&
\left. {\partial_r X\over\partial J_B}\right|_K
\left. {\partial_r J_B\over \partial \Phi^A}\right|_K
\left. {\partial_l Y\over\partial K_A}\right|_\Phi-
\left. {\partial_r X\over\partial K_A}\right|_\Phi
\left. {\partial_l Y\over\partial \Phi^A}\right|_K
\nonumber\\
&=&\left.{\partial_r X\over \partial J_B}\right|_K(J_B,Y)
-\left. {\partial_rX\over\partial K_A}\right|_J
\left. {\partial_l J_A\over\partial\Phi^B}\right|_K
\left.{\partial_l Y\over \partial J_B}\right|_K(-1)^{\varepsilon_A+
\varepsilon_B+\varepsilon_A\varepsilon_B}.
\end{eqnarray}
Now, notice that the first term of the above expression can
be written as
\begin{eqnarray}
\left.{\partial_r X\over \partial J_B}\right|_K(J_B,Y)
&=&-(-1)^{\varepsilon_B}\left.{\partial_r X\over \partial J_B}\right|_K
\left. {\partial_l (\Gamma,Y)\over\partial\Phi^B}\right|_K+
\left.{\partial_r X\over \partial J_B}\right|_K
\left(\Gamma, \left. {\partial_l Y\over\partial\Phi^B}\right|_K\right)
\nonumber\\
&=&-\{X,(\Gamma,Y)\}+
\left.{\partial_r X\over \partial J_B}\right|_K
\left(\Gamma, \left. {\partial_l Y\over\partial\Phi^B}\right|_K\right).
\label{y1}
\end{eqnarray}
On the other hand, the second term is
\begin{eqnarray}
\phantom{.}&&
-\left. {\partial_rX\over\partial K_A}\right|_J
\left. {\partial_l J_A\over\partial\Phi^B}\right|_K
\left.{\partial_l Y\over \partial J_B}\right|_K(-1)^{\varepsilon_A+
\varepsilon_B+\varepsilon_A\varepsilon_B}\nonumber\\&&
=-(-1)^{\varepsilon_A+\varepsilon_B\varepsilon(X)}\left\{
\left.{\partial_l\over\partial\Phi^B}\left(
\left.{\partial_r X\over\partial K_A}\right|_JJ_A\right)\right|_K-
\left.{\partial_l\over\partial\Phi^B}\left(
\left.{\partial_r X\over\partial K_A}\right|_J\right)\right|_KJ_A\right\}
\left.{\partial_l Y\over\partial J_B}\right|_K.
\end{eqnarray}
Remembering that \cite{me}
\begin{equation}
(-1)^{\varepsilon_A}\left.{\partial_r Z\over \partial K_A}\right|_J J_A=
(Z,\Gamma)\quad\quad\forall Z,
\end{equation}
we get
\begin{eqnarray}
\phantom{.}&&-\left. {\partial_rX\over\partial K_A}\right|_J
\left. {\partial_l J_A\over\partial\Phi^B}\right|_K
\left.{\partial_l Y\over \partial J_B}\right|_K(-1)^{\varepsilon_A+
\varepsilon_B+\varepsilon_A\varepsilon_B}
=-(-1)^{\varepsilon_B(\varepsilon(X)+\varepsilon(Y)+1)}
\left.{\partial_l (X,\Gamma)\over \partial \Phi^B}\right|_K
\left.{\partial_r Y\over \partial J_B}\right|_K\nonumber\\&&
+(-1)^{\varepsilon(X)\varepsilon_B}
\left.{\partial_l J_C\over\partial\Phi^B}\right|_K\left(
\left.{\partial_l X\over\partial J_C}\right|_K,\Gamma\right)
\left. {\partial_l Y\over\partial J_B}\right|_K.
\label{y2}
\end{eqnarray}
At this point, collecting (\ref{y1}) and (\ref{y2}),
it is simple to arrive at the desired result,
i.e.\ formula (\ref{identity}).

It is, instead, immediate to prove that
\begin{equation}
dX=d_J X-\{d\omega_0,X\}=d_JX-\{(\omega_1,\omega_0),X\},
\label{dij}
\end{equation}
where $d_J$ differs from $d$ for the fact that $\{J,K\}$ are kept
constant instead of $\{\Phi,K\}$.

We are now ready to prove (\ref{deomega1}). Using (\ref{dij}) with
$X=\omega_1$, we get
\begin{equation}
d\omega_1=d_J\omega_1-\{(\omega_1,\omega_0),\omega_1\}.
\label{eq0}
\end{equation}
On the other hand, (\ref{identity}) with $X=Y=\omega_1$ gives
\begin{equation}
\{(\omega_1,\omega_0),\omega_1\}=-{1\over 2}(\omega_1,\omega_1)-{1\over 2}
(\omega_0,\{\omega_1,\omega_1\}).
\label{eq1}
\end{equation}
Moreover, a straightforward differentiation permits to write
\begin{equation}
d_J\omega_1=d_J<\chi>_J=<d\chi>_J-{i\over \hbar}<\chi \, d\Sigma>_J
+{i\over \hbar}<\chi>_J<d\Sigma>_J.
\end{equation}
Using (\ref{dechi}) and $d\Sigma=\Omega\chi$
[which is the second of (\ref{usefulid})], we obtain
\begin{equation}
d_J\omega_1=-{i\over 2\hbar}<\Omega(\chi\chi)>_J+{i\over \hbar}
\omega_1(\omega_1,\omega_0)=-{i\over 2\hbar}
(<\chi\chi>_J-\omega_1\omega_1,\omega_0).
\label{eq2}
\end{equation}
Collecting (\ref{eq0}), (\ref{eq1}) and (\ref{eq2}) we arrive directly at
(\ref{deomega1}).

Now, let us come to the diagrammatical meaning of the expression
$\omega_2$ given in formula (\ref{omega2}). The term
\begin{equation}
{i\over 2\hbar}<\chi\chi>_J
\label{t1}
\end{equation}
collects a set of Feynmann
diagrams that contain two insertions of the composite
operator $\chi$. These diagrams are connected and irreducible as for
the action vertices, but are neither connected nor
irreducible as for the vertices that represent the
$\chi$-insertions. Thus, it is natural to conjecture that $\omega_2$
represents the set of connected irreducible graphs with two $\chi$ insertions,
i.e.\ that the remaining terms of expression (\ref{omega2})
remove the disconnected and reducible contributions
that are contained in (\ref{t1}). This is indeed true.
The term
\begin{equation}
-{i\over 2\hbar}\omega_1\omega_1
\end{equation}
subtracts the disconnected contributions, i.e.\ all the
graphs that are a product of two separate diagrams, each one containing
a single $\chi$-insertion. On the other hand, the term
\begin{equation}
-{1\over 2}\{\omega_1,\omega_1\}=-{1\over 2}
\left.{\partial \omega_1\over \partial\Phi^A}\right|_K
\left. {\partial _l^2 W[J,K]\over \partial J_A\partial J_B}\right|_K
\left.{\partial \omega_1\over \partial\Phi^B}\right|_K
\end{equation}
represents the set of connected reducible diagrams, i.e.\
those diagrams in which a single leg connects the two $\chi$-insertions.
Indeed, the derivatives with respect to $\Phi$ represent $\Phi$-legs,
while $\left. {\partial _l^2 W[J,K]\over \partial J_A\partial J_B}\right|_K$
is the propagator connecting them. As promised, this also provides a nice
diagrammatical interpretation of the mixed brackets.

The diagrammatical meaning of $\omega_2$ is crucial, in the sense that it
guarantees that the overall divergences of $\omega_2$ are local, when
all subdivergences have been removed. This fact will be important is
the sequel.

Identity (\ref{deomega1}) can be rewritten as
\begin{equation}
{\cal F}=d\omega_1-{1\over 2}(\omega_1,\omega_1)=
-{\rm ad}\, \Gamma \,\,\omega_2.
\label{fieldstrength}
\end{equation}
Its meaning is that the field strength ${\cal F}$ of the
connection $\omega_1$ does not vanish,
however it is ${\rm ad}\, \Gamma$-exact (and so, it vanishes on shell).
Notice that, due to (\ref{covder}), the field strength
of the connection $d\Psi$ is instead zero.
Similarly, due to (\ref{dechi}), the field strength of $\chi$ is also zero.
However, as we shall discuss later on, the latter fact is related to the
special form of $\chi$ that we have chosen. Indeed, if we let
$\chi$ go into
$\chi^\prime=\chi+\Omega \delta$, $\delta$ being some local functional,
[this is allowed, because it
does not affect the second equation of (\ref{usefulid}) that defines
$\chi$],
then the field strength of $\chi^\prime$ is no more zero, but it is
$\Omega$-exact.

We can collect equations (\ref{deomega1}), (\ref{uopo}) and the
Ward identity
$(\Gamma,\Gamma)=0$ into the formula
\begin{equation}
d\omega_i={1\over 2}(-1)^{j+1}(\omega_j,\omega_{i-j+1}),
\label{deomegai}
\end{equation}
where $i,j=-1,0,1$ and the sum over $j$ is understood.
We therefore are lead to conjecture that there exist $k$-forms
$\omega_k$ on ${\cal M}_{gf}$, for $k=3,\ldots m$,
such that (\ref{deomegai})
holds for $i,j=-1,\ldots m$.
It is also natural to conjecture that the $k$-form
$\omega_k$, $k=3,\ldots m$ represents
the connected irreducible Feynman diagrams with $k$ $\chi$-insertions.
Equations (\ref{deomegai}) will be called
{\sl cascade equations} due to their aspect and
are a generalization of the descent equations
that would read, in this case,
\begin{equation}
d\omega_i={\rm ad} \,\Gamma\,\, \omega_{i+1}.
\end{equation}
Such a formula is correct only for $i=-1,0$. All the other cases are improved
by the fact that the exterior derivative $d$ and the
nilpotent operator ${\rm ad}\, \Gamma$ do not commute nor anticommute,
rather
\begin{equation}
[d,{\rm ad}\, \Gamma]X=(-1)^{d_X}(X,d\Gamma)
\end{equation}
(the square brackets still denote a commutator).
On the other hand, the operator ${\rm ad}\, \Gamma$ and the covariant
derivative ${\cal D}=d-{\rm ad}_l\, \omega_1$ have the following properties
\begin{equation}
\matrix{
({\rm ad}\, \Gamma)^2=0,&\quad [{\cal D},{\rm ad}\, \Gamma]=0,&
\quad {\cal D}^2=({\rm ad}\, \Gamma \,\, {\sl something},\, . \, ),}
\label{12}
\end{equation}
so that ${\cal D}^2=0$ on shell. This is the difference with respect to
a double-complex and is responsible of the generation of cascade
equations instead of descent equations. A structure like (\ref{12})
will be called a {\sl quasi} double-complex.

Let us discuss the properties of the cascade equations (\ref{deomegai}),
beginning from self-consistency. Interchanging the forms in the right-hand
side we get
\begin{equation}
d\omega_i={1\over 2}(-1)^{i-j}(\omega_{i-j+1},\omega_i),
\end{equation}
which is consistent with the replacement $j\rightarrow i-j+1$.
Moreover, taking the exterior derivative and using eq.s (\ref{deomegai})
themselves, we get
\begin{equation}
(\omega_{i-j+1},(\omega_k,\omega_{j-k+1}))(-1)^k=0,
\end{equation}
which is indeed true (it is sufficient to use the Jacobi identity for the
antibrackets and help oneself with replacements of indices).

The cascade equations possess some ``invariance", similar to the
invariance under $\omega_i\rightarrow \omega_i+\sigma \Delta_i+
d\Delta_{i-1}$ of the usual descent equations,
however much more complicated. Let us
consider the replacement $\omega_1\rightarrow\omega_1^\prime
=\omega_1+(\Delta_1,\omega_0)$. One checks that
there is invariance under
\begin{eqnarray}
\omega_0^\prime&=&\omega_0,\nonumber\\
\omega_1^\prime&=&\omega_1+(\Delta_1,\omega_0),\nonumber\\
\omega_2^\prime&=&\omega_2-d\Delta_1+(\Delta_1,\omega_1)
+{1\over 2!}(\Delta_1,(\Delta_1,\omega_0)),\nonumber\\
\omega_3^\prime&=&\omega_3-{1\over 2!}(\Delta_1,d\Delta_1)+
(\Delta_1,\omega_2)+
{1\over 2!}(\Delta_1,(\Delta_1,\omega_1))
+{1\over 3!}(\Delta_1,(\Delta_1,(\Delta_1,\omega_0))),\nonumber\\
&\ldots &
\label{invariance}
\end{eqnarray}
Similar formul\ae\ hold starting from every $\omega_i$ with
$\omega_i\rightarrow
\omega_i+(\Delta_i,\omega_0)$, $i=-1,\ldots m$.
For example, there is invariance under
\begin{eqnarray}
\omega_0^\prime&=&\omega_0,\nonumber\\
\omega_1^\prime&=&\omega_1,\nonumber\\
\omega_2^\prime&=&\omega_2+(\Delta_2,\omega_0),\nonumber\\
\omega_3^\prime&=&\omega_3+d\Delta_2+(\Delta_2,\omega_1),\nonumber\\
\omega_4^\prime&=&\omega_4+(\Delta_2,\omega_2)+
{1\over 2!}(\Delta_2,(\Delta_2,\omega_0)),\nonumber\\
\omega_5^\prime&=&\omega_5+
{1\over 2!}(\Delta_2,d\Delta_2)+
(\Delta_2,\omega_3)+
{1\over 2!}(\Delta_2,(\Delta_2,\omega_1)),\nonumber\\
&\ldots &
\label{invariance2}
\end{eqnarray}
The proof of invariance under (\ref{invariance}) and (\ref{invariance2})
is straightforward. Again, due to their aspect, invariances (\ref{invariance}),
(\ref{invariance2}) and similar ones will be called {\sl cascade invariances}
or {\sl cascade transformations}. The cascade transformation that starts from
$\omega_i\rightarrow \omega_i+(\Delta_i,\omega_0)$ will
be called {\sl cascade transformation of degree} $i$.

Let us discuss a particular case of the above transformations. Let us consider
a cascade transformation of degree one (\ref{invariance}) with
\begin{equation}
\Delta_1=<\delta_1>_J,
\end{equation}
$\delta_1$ being a local functional of fields and BRS sources. Since
$\omega_1=<\chi>_J$, (\ref{invariance}) gives
\begin{equation}
\omega_1^\prime=<\chi^\prime>_J\equiv <\chi+\Omega \delta_1>_J.
\end{equation}
Thus, it is natural to expect that $\chi\rightarrow\chi^\prime$
generates the same transformation as (\ref{invariance}). We now show that
it is not precisely so, however $\chi\rightarrow\chi^\prime$ generates
(\ref{invariance}) up to a cascade transformation of degree two, eq.s
(\ref{invariance2}),
$\Delta_2$ being a suitable set of connected irreducible diagrams
with two insertions of certain local composite operators.

Let us first notice that equation (\ref{dechi}) is modified into
\begin{equation}
d\chi^\prime={1\over 2}(\chi^\prime,\chi^\prime)+
\Omega\left\{d\delta_1-(\delta_1,\chi)-{1\over 2}
(\delta_1,\Omega\delta_1)\right\}.
\label{dechiprime}
\end{equation}
The transformed field strength is no more zero, however it is
$\Omega$-exact [we anticipated this fact soon after
formula (\ref{fieldstrength})].
Repeating the argument of (\ref{eq0}), (\ref{eq1}) and
(\ref{eq2}) with the primed functionals and
using the appropriate expression for $d\chi^\prime$ given
in (\ref{dechiprime}), one finds
\begin{eqnarray}
\omega_2^\prime&=&{i\over 2\hbar}<\chi^\prime\chi^\prime>_J-
{i\over2\hbar}\omega_1^\prime\omega_1^\prime
-{1\over 2}\{\omega_1^\prime,\omega_1^\prime\}\nonumber\\&&
-<d\delta_1>_J+
<(\delta_1,\chi)>_J+{1\over 2!}<(\delta_1,\Omega\delta_1)>_J.
\label{omega2prime}
\end{eqnarray}
This form of $\omega_2^\prime$ differs from the one given
in (\ref{invariance}) and
it is clearly the sum of connected irreducible Feynman graphs.
The difference between the
expressions for $\omega_2^\prime$ given in (\ref{invariance}) and
in (\ref{omega2prime}) is easily shown to be equal to
\begin{equation}
({\cal U}_1+{\cal U}_2,\omega_0),
\label{uopo2}
\end{equation}
where
\begin{eqnarray}
{\cal U}_1&=&{i\over \hbar}(<\delta_1\chi>_J-\Delta_1\omega_1)
-\{\Delta_1,\omega_1\},\nonumber\\
{\cal U}_2&=&-{i\over 2\hbar}\left(<\Omega\delta_1 \,\delta_1>_J-
(\Delta_1,\omega_0)\Delta_1\right)
+{1\over 2}\{(\Delta_1,\omega_0),\Delta_1\}.
\end{eqnarray}
Now that we know how to express the set of connected irreducible graphs
with two insertions of local composite operators [see (\ref{omega2})],
it is evident that both ${\cal U}_1$ and ${\cal U}_2$ represent such
situations.
We conclude that the two expressions for $\omega_2^\prime$ differ by
a transformation of the kind (\ref{invariance2}) with
\begin{equation}
\Delta_2={\cal U}_1+{\cal U}_2.
\end{equation}
$\Delta_2$, expression (\ref{uopo2}) and both the expressions of
$\omega_2^\prime$ of eq.\ (\ref{invariance}) and eq.\ (\ref{omega2prime}),
have the property that their order $\hbar^n$ divergences are local, when
all subdivergences have been removed.

This discussion illustrates that the cascade transformations of the
kind $\Delta_i=<\delta_i>_J$ do not spoil the property that
the ${\cal O}(\hbar^n)$ of $\omega_k$ are local when all the
subdivergences have been removed. In particular, notice that if $X$
and $Y$ are sums of connected irreducible diagrams, then $(X,Y)$
has the same property. Indeed, antibrackets connect a $K_A$-leg
of $X$ with a $\Phi^A$-leg of $Y$, and viceversa, {\sl without}
any propagator on the $\Phi^A$- and $K_A$-legs. This defines new
vertices obtained by shrinking the $\Phi^A$-$K_A$ leg to a point
and the graphs of $(X,Y)$
constructed with such vertices are irreducible.

The cascade equations are also invariant
under canonical transformations.
Let $ F (\Phi,K^\prime)$ denote the generating functional.
Then one has
\begin{equation}
\matrix{\omega_0^\prime=\tilde\omega_0,&\quad
\omega_1^\prime=\tilde\omega_1-d^\prime  F ,&\quad
\omega_2^\prime=\tilde\omega_2,&\quad
\omega_3^\prime=\tilde\omega_3,&\quad\ldots}
\label{caninvariance}
\end{equation}
where $d^\prime$ is a derivative at constant $\{\Phi,K^\prime\}$.
As we see, only $\omega_1$ has a strange transformation rule,
which is nothing
but the analogue of (\ref{chitr}) and is very similar to a gauge
transformation. It will be further investigated in section
\ref{application}. (\ref{caninvariance}) can
be proved starting from
the properties
\begin{eqnarray}
d\tilde X&=&\widetilde{d X}-(d^\prime  F ,\tilde X),\nonumber\\
dd^\prime  F &=&-{1\over 2}(d^\prime  F , d^\prime  F ).
\end{eqnarray}
The first equation is analogous to the formula that was proved in the
appendix of ref.\ \cite{me} and can be derived following similar
steps. The second equation is derived as follows. We write
\begin{eqnarray}
dd^\prime  F &=&d\kappa_i\left.{\partial d^\prime  F
\over \partial\kappa_i}
\right|_{\Phi,K}\nonumber\\&=&
d\kappa_i\left.{\partial d^\prime  F \over \partial\kappa_i}
\right|_{\Phi,K^\prime}+
d\kappa_i\left.{\partial K^\prime_A\over \partial\kappa_i}
\right|_{\Phi,K}\left.{\partial d^\prime  F \over \partial K^\prime_A}
\right|_{\Phi}\nonumber\\&=&
d^\prime d^\prime  F
-\left.{\partial d^\prime  F \over \partial\Phi^{\prime A}}
\right|_{K^\prime}\left.{\partial d^\prime  F \over \partial K^\prime_A}
\right|_{\Phi}\nonumber\\&=&
-\left.{\partial d^\prime  F \over \partial\Phi^{\prime A}}
\right|_{K^\prime}\left.{\partial d^\prime  F \over \partial K^\prime_A}
\right|_{\Phi^\prime}-\left.{\partial d^\prime  F \over
\partial\Phi^{\prime A}}
\right|_{K^\prime}F^{BA}\left.{\partial d^\prime  F
\over \partial \Phi^{\prime B}}
\right|_{K^\prime}.
\end{eqnarray}
The first term in the last expression is equal to $-{1\over 2}
(d^\prime  F ,d^\prime  F )$, while the second term (where
$F^{BA}={\partial_l^2 F\over \partial K^\prime_A\partial K^\prime_B}$)
vanishes. This can be easily proved by interchanging the various factors and
showing that the expression is equal to the opposite of itself.

The same formul\ae\ prove that the field strength ${\cal F}$
of (\ref{fieldstrength})
is sent into $\tilde{\cal F}$, a fact that has a nice interpretation:
the canonical
transformations are the gauge-transformations for the connection $\omega_1$.
A completely analogous argument
proves that (\ref{dechi}) is canonically preserved,
a property that we left without proof before.
Indeed, the transformation rule for $\chi$ (\ref{chitr})
is formally analogous to that of
$\omega_1$.

Equations (\ref{deomegai}) where rigorously derived for $i=-1,0,1$ and then
conjectured for the other values of $i$. Although they seem very natural
(and their properties together with the applications that will be
examined in the next section give a stronger support to this), we want to
conclude this section with the
explicit proof of the case $i=2$.
This provides an expression for $\omega_3$
that permits to interpret it as the set of connected irreducible Feynman
diagrams with three $\chi$-insertions. Moreover, the explicit calculation
of $d\omega_2$ is sufficient to illustrate how to proceed in
the proofs of any case $i>2$ of (\ref{deomegai}).
Let us start by applying equations (\ref{dij}) and (\ref{identity}), obtaining
\begin{equation}
d\omega_2=(\omega_1,\omega_2)-(\{\omega_1,\omega_2\},\omega_0)+
d_J\omega_2+\{\omega_1,(\omega_2,\omega_0)\}.
\end{equation}
A direct computation and a series of manipulations that by now should have
become standard give
\begin{eqnarray}
d_J\omega_2&=&-{1\over 3!\hbar^2}
(<\chi\chi\chi>,\omega_0)+{1\over 2\hbar^2}<\chi\chi>
(\omega_1,\omega_0)+
\{\omega_1,d\omega_1\}-{i\over \hbar}\{d\Gamma,\omega_1\}\omega_1
\nonumber\\&&
-{i\over 2\hbar}
(\omega_1,\omega_1)\omega_1
+{i\over \hbar}(\omega_2,\omega_0)\omega_1
-{1\over 2}\left\{d\Gamma,\left.
{\partial_r\omega_1\over\partial\Phi^A}
\right|_K\right\}\left.{\partial_l\omega_1\over \partial J_A}\right|_K
\nonumber\\&&+{1\over 2}\{\omega_1,\{d\Gamma,\omega_1\}\}.
\end{eqnarray}
Collecting the above two formul\ae\ and
using (\ref{omega2}), we arrive at
\begin{eqnarray}
d\omega_2-(\omega_1,\omega_2)+(\Delta\omega_3,\omega_0)&=&
{1\over 2}\{\omega_1,(\omega_1,\omega_1)\}+{1\over 2}\{
\omega_1,\{d\Gamma,\omega_1\}\}\nonumber\\&&
-{1\over 2}
\left\{d\Gamma,\left.{\partial_r\omega_1\over\partial\Phi^A}
\right|_K\right\}\left.{\partial_l\omega_1\over \partial J_A}\right|_K,
\label{partial}
\end{eqnarray}
where $\Delta\omega_3$ is the first set of contributions to $\omega_3$,
precisely
\begin{equation}
\Delta\omega_3={1\over 3!\hbar^2}
<\chi\chi\chi>-{1\over 3!\hbar^2}\omega_1\omega_1\omega_1
+\{\omega_1,\omega_2\}+{i\over \hbar}
\omega_2\omega_1+
{i\over 2\hbar}\{\omega_1,\omega_1\}\omega_1.
\end{equation}
It is not difficult to prove that
\begin{equation}
-{1\over 2}
\left\{d\Gamma,\left.{\partial_r\omega_1\over\partial\Phi^A}
\right|_K\right\}\left.{\partial_l\omega_1\over \partial J_A}\right|_K
=-{1\over 4}\{d\Gamma,\{\omega_1,\omega_1\}\}
+{1\over 4}\{d\Gamma,\omega_1,\omega_1\},
\label{partial2}
\end{equation}
where we have introduced the following notion of {\sl triple brackets}
\begin{equation}
\{X,Y,Z\}=
\left.{\partial_r X\over \partial\Phi^A}\right|_K
\left.{\partial_l^3 W\over \partial J_A\partial J_B \partial J_C}\right|_K
\left.{\partial_l Y\over \partial\Phi^C}\right|_K
\left.{\partial_r Z\over \partial\Phi^B}\right|_K
(-1)^{\varepsilon_B(\varepsilon(Y)+\varepsilon(Z)+1)}.
\end{equation}
The triple brackets have the following properties
\begin{eqnarray}
\{Y,X,Z\}&=&(-1)^{\varepsilon(X)\varepsilon(Y)+d_Xd_Y}
\{X,Y,Z\}\nonumber\\
\{X,Z,Y\}&=&(-1)^{\varepsilon(Y)\varepsilon(Z)+d_Yd_Z}
\{X,Y,Z\}.
\end{eqnarray}
The diagrammatical meaning of the triple brackets is that, when
$X$, $Y$ and $Z$ are average values of local composite operators,
$\{X,Y,Z\}$ is the set of connected reducible diagrams where the insertions
of the three composite operators are only connected {\sl via}
the three-vertex
\begin{equation}
\left.{\partial_l^3 W\over \partial J_A\partial J_B \partial J_C}\right|_K.
\end{equation}
This will be illustrated in more detail later on.
An identity generalizing (\ref{identity}) for $(\{X,Y,Z\},\omega_0)$
can be surely proved for the triple
brackets, however we only need the case $X=Y=Z=\omega_1$,
in which it happens that
\begin{equation}
{1\over 3}(\{\omega_1,\omega_1,\omega_1\},\omega_0)=
\{\omega_1,(\omega_1,\omega_1)\}+(\omega_1,\{\omega_1,\omega_1\})
+\{d\Gamma,\omega_1,\omega_1\}.
\label{identity2}
\end{equation}
Indeed, one has
\begin{equation}
(\omega_1,\{\omega_1,\omega_1\})+
\{\omega_1,(\omega_1,\omega_1)\}=
(-1)^{\varepsilon_A(1+\varepsilon_B)}\left(\omega_1,
\left.{\partial_r^2 \Gamma\over \partial\Phi^A\partial\Phi^B}\right|_K\right)
\left.{\partial\omega_1\over\partial J_B}\right|_K
\left.{\partial\omega_1\over\partial J_A}\right|_K.
\end{equation}
Moreover,
\begin{eqnarray}
(\{\omega_1,\omega_1,\omega_1\},\Gamma)&=&J_A
\left.{\partial_r \{\omega_1,\omega_1,\omega_1\}\over \partial K_A}
\right|_J
=3\{d\Gamma,\omega_1,\omega_1\}\nonumber\\&&
-3(-1)^{(\varepsilon_A+\varepsilon_C)(1+\varepsilon_B)+
\varepsilon_A\varepsilon_C}
\left. {\partial \omega_1\over \partial K_C}\right|_J
\left.{\partial_r^3 \Gamma\over \partial \Phi^C
\partial\Phi^A\partial\Phi^B}\right|_K
\left.{\partial\omega_1\over\partial J_B}\right|_K
\left.{\partial\omega_1\over\partial J_A}\right|_K\nonumber\\&&
+(-1)^{\varepsilon_B}
\left.{\partial\omega_1\over\partial J_A}\right|_K
\left.{\partial\omega_1\over\partial J_B}\right|_K
\left.{\partial\omega_1\over\partial J_C}\right|_K
\left(\left.{\partial_r^3 \Gamma\over \partial
\Phi^A\partial\Phi^B\partial\Phi^C}\right|_K,
\Gamma\right)
\end{eqnarray}
Differentiating $(\Gamma,\Gamma)=0$ three times with respect to $\Phi$,
one finds
\begin{eqnarray}
\phantom{.}&&(-1)^{\varepsilon_B}
\left(\left.{\partial_r^3 \Gamma\over \partial \Phi^A\partial\Phi^B
\partial\Phi^C}\right|_K,
\Gamma\right)=-(-1)^{\varepsilon_A+\varepsilon_B}
\left(
\left.{\partial_r^2 \Gamma\over \partial\Phi^B\partial\Phi^C}\right|_K,
\left.{\partial_r \Gamma\over \partial \Phi^A}\right|_K\right)
\nonumber\\&&
-(-1)^{\varepsilon_A}
\left(
\left.{\partial_r \Gamma\over \partial\Phi^C}\right|_K,
\left.{\partial_r^2 \Gamma\over \partial \Phi^A\partial\Phi^B}\right|_K\right)
-(-1)^{\varepsilon_A\varepsilon_B}
\left(
\left.{\partial_r^2 \Gamma\over \partial\Phi^A\partial\Phi^C}\right|_K,
\left.{\partial_r \Gamma\over \partial \Phi^B}\right|_K\right).
\end{eqnarray}
Using this equation, one finally arrives at (\ref{identity2}).
Collecting (\ref{partial}), (\ref{partial2}) and (\ref{identity2}), one finally
gets the desired equation, namely
\begin{equation}
d\omega_2=(\omega_1,\omega_2)+(\omega_3,\omega_0),
\end{equation}
where
\begin{eqnarray}
\omega_3&=&{1\over 3!\hbar^2}(
<\chi\chi\chi>-\omega_1\omega_1\omega_1)
+\{\omega_1,\omega_2\}+{i\over \hbar}
\omega_2\omega_1+
{i\over 2\hbar}\{\omega_1,\omega_1\}\omega_1\nonumber\\&&-
{1\over 4}\{\omega_1,\{\omega_1,\omega_1\}\}
+{1\over 12}\{\omega_1,\omega_1,\omega_1\}.
\end{eqnarray}
The diagrammatical meaning of each term is illustrated in Fig.\ 1. The blobs
denote sets of connected irreducible diagrams, the dots denote the
$\chi$-insertions. It is a useful exercise to check that each coefficient
and each power of $\hbar$ is in agreement with the diagrammatical meaning
of $\omega_3$.

\begin{figure}
\begin{center}
\caption{$\omega_3$ is a sum
of connected irreducible Feynmann diagrams.}
\begin{picture}(220,297.5)(0,-17.5)
\thicklines

\put(110,230){\circle{40}}
\put(110,213){\circle*{2}}
\put(125,239){\circle*{2}}
\put(95,239){\circle*{2}}

\put(205,240){\circle{15}}
\put(175,240){\circle{15}}
\put(190,212){\circle{15}}
\put(190,212){\circle*{2}}
\put(205,240){\circle*{2}}
\put(175,240){\circle*{2}}

\put(45,140){\oval(15,15)[r]}
\put(15,140){\oval(15,15)[l]}
\put(30,112){\circle{15}}
\put(15,147.5){\line(1,0){30}}
\put(15,132.5){\line(1,0){30}}
\put(30,132.5){\line(0,-1){13}}
\put(30,112){\circle*{2}}
\put(45,140){\circle*{2}}
\put(15,140){\circle*{2}}

\put(125,140){\oval(15,15)[r]}
\put(95,140){\oval(15,15)[l]}
\put(110,112){\circle{15}}
\put(95,147.5){\line(1,0){30}}
\put(95,132.5){\line(1,0){30}}
\put(110,112){\circle*{2}}
\put(125,140){\circle*{2}}
\put(95,140){\circle*{2}}

\put(205,140){\circle{15}}
\put(175,140){\circle{15}}
\put(190,112){\circle{15}}
\put(182.5,140){\line(1,0){15}}
\put(190,112){\circle*{2}}
\put(205,140){\circle*{2}}
\put(175,140){\circle*{2}}

\put(70,40){\circle{15}}
\put(40,40){\circle{15}}
\put(55,12){\circle{15}}
\put(47.5,40){\line(1,0){15}}
\put(55,12){\circle*{2}}
\put(70,40){\circle*{2}}
\put(40,40){\circle*{2}}
{\thinlines
\put(70,40){\oval(19,19)[r]}
\put(40,40){\oval(19,19)[l]}
\put(40,49.5){\line(1,0){30}}
\put(40,30.5){\line(1,0){30}}
}
\put(55,30.5){\line(0,-1){11}}

\put(184.5,44.5){\circle{15}}
\put(145.5,44.5){\circle{15}}
\put(165,7.5){\circle{15}}
\put(165,7.5){\circle*{2}}
\put(184.5,44.5){\circle*{2}}
\put(145.5,44.5){\circle*{2}}
\put(165,30){\circle{8}}
\put(165,15){\line(0,1){11}}
\put(168.5,32.5){\line(3,2){10}}
\put(161.5,32.5){\line(-3,2){10}}

\put(30,230){\makebox(0,0){${1\over 3!\hbar^2}<\chi\chi\chi>_J=$}}
\put(110,190){\makebox(0,0){$\omega_3$}}
\put(190,190){\makebox(0,0){$+{1\over 3! \hbar^2}
\omega_1\omega_1\omega_1$}}
\put(30,90){\makebox(0,0){$-\{\omega_1,\omega_2\}$}}
\put(110,90){\makebox(0,0){$-{i\over\hbar}\omega_2\omega_1$}}
\put(190,90){\makebox(0,0){$-{i\over 2!\hbar}
\{\omega_1,\omega_1\}\omega_1$}}
\put(55,-10){\makebox(0,0){$+{1\over 4}\{\omega_1,
\{\omega_1,\omega_1\}\}$}}
\put(165,-14.5){\makebox(0,0){$-{1\over12}\{\omega_1,
\omega_1,\omega_1\}$}}

\end{picture}
\end{center}
\end{figure}

\section{Application of the covariant formalism}
\label{application}
\setcounter{equation}{0}

It is now time to apply the formalism developed
in the previous section in order to find a functional $R^{(n)}$
that makes every $\omega_k$ finite up to order $\hbar^n$, under the
inductive hypothesis that $\omega_k$ are finite up to order $\hbar^{n-1}$.
This inductive assumption (that suitably extends
the one that was needed for a single gauge-fixing parameter $\kappa$)
is clearly satisfied for $n=0$: indeed, for
$n=0$ (i.e.\ for the theory described by the action $\Sigma_0$),
the order $\hbar^0$ of
$\omega_0$ is $S$, which is
clearly finite, the order $\hbar^0$ of $\omega_1$ is $d\Psi$,
which is also finite,
while the order $\hbar^0$ parts of $\omega_k$ for $k>1$
are zero.
The reason of this last fact is that for $k>1$ the number of $\chi$-insertions
is greater that one and of course
there is no connected irreducible tree diagram with more that
one $\chi$-insertion.

Let us recall what is the situation. We possess
cascade equations for the ``$n-1$-theory'', namely the theory described by
$\Sigma_{n-1}$, that is convergent up to order $\hbar^{n-1}$ included.
The forms
$\omega_k^{(n-1)}$ are finite up to
order $\hbar^{n-1}$, by induction.
We want to make them finite up to order $\hbar^n$.
The effective action $\Gamma_{n-1}=\omega_0^{(n-1)}$ is promptly
made finite up to order $\hbar^n$ by suitably redefining the
parameters $\lambda_i$ and by performing a canonical transformation
generated by (\ref{fn}), as explained in section
\ref{generalization}. This only changes $\omega_k^{(n-1)}$
up to order $\hbar^n$,
so that the inductive assumption is preserved: $\omega_k^{(n-1)}$
is finite up to order $\hbar^{n-1}$ $\forall k$
and moreover $\omega_0^{(n-1)}$ is turned into
$\Gamma_n=\omega_0^{(n)}$, which is finite
up to order $\hbar^n$. This is sufficient, as in
section \ref{generalization}, to show that
\begin{equation}
d\delta_n\lambda_i=0\hskip .3truecm \forall i,
\label{41}
\end{equation}
since the argument of section
\ref{generalization} is trivially extended to an arbitrary number of
gauge-fixing parameters $\kappa$.
(\ref{41}) permits to derive cascade equations of the ``$n$-theory'',
i.e.\ for the forms $\omega_k^{(n)}$ [from now on,
we suppress the superscript $(n)$]. Clearly, we can choose
$\omega_k$ such that $\omega_k=\omega_k^{(n-1)}+
{\cal O}(\hbar^n)={\rm finite}+{\cal O}(\hbar^n)$.
Thus, we remain with the task of showing how to make
$\omega_k$ finite
up to order $\hbar^n$ for $k\geq 1$ in order to fully reproduce
the inductive assumption to order $\hbar^n$.
The argument given at the end of section
\ref{generalization} for
a single gauge-fixing parameter
is not immediately generalizable to the case of
many gauge-fixing parameters, as anticipated.
We need to combine the properties
of the cascade equations together with the arbitrariness (\ref{freedom}).

Before doing this, we need to discuss the implications of the cascade equations
for $\omega_k$ and the properties satisfied by their order $\hbar^n$
divergent parts $\omega_{k\, div}$. First of all, we notice that
the cascade equations for $\omega_k$
imply descent equations for $\omega_{k\, div}$.
The quasi double-complex generates a double complex.
It is sufficient to take the order $\hbar^n$
divergent parts of eq.s (\ref{deomegai}). We get
\begin{equation}
D\omega_{k\, div}=(-1)^k\sigma\omega_{k+1\, div}.
\label{descent}
\end{equation}
These equations are indeed descent equations, since
$D^2=0$, $D\sigma=\sigma D$ and $\sigma^2=0$.

Choosing the functionals $\Delta_k={\cal O}(\hbar^n)$,
$k=1,\ldots m$, to be averages of local functionals,
the cascade invariances reduce to the usual invariances of the descent
equations, namely
\begin{equation}
\omega_{k\, div}^\prime=\omega_{k\, div}-(-1)^kD\Delta_{k-1\, div}
+\sigma\Delta_{k\, div},
\label{cascadediv}
\end{equation}
where $\Delta_{k\, div}$ denotes the (local) order $\hbar^n$ divergent
part of $\Delta_k$ and we have set $\Delta_{-1}=\Delta_0=0$.

The fact that the covariant derivative $D$ is flat, $D^2=0$,
permit to ``integrate'' a $D$-closed form $\omega$, $D\omega=0$,
i.e.\ to find (perturbatively in the gauge-fixing parameters
$\{\kappa_1,\ldots,\kappa_m\}$)
a form $\delta$ such that $\omega=D\delta$. An example of this
kind was given at the end of section \ref{generalization},
where (\ref{eqa}) solved (\ref{eqb}). Analogous formul\ae\ can be
found for a generic $\omega$, using the ``integrability condition''
$D\omega=0$. The general solution is obtained as follows.
Let us introduce the vector field
\begin{equation}
v=\kappa_i{\partial\over \partial \kappa_i}
\end{equation}
and let $i_v$ denote the operator of contraction with $v$.
Furthermore, we need the following operator
\begin{equation}
\Theta=\kappa_i D_i.
\end{equation}
It is easy to
prove the following properties
\begin{equation}
\matrix{
[D,i_v]_+\omega_p=(p+\Theta)\omega_p,&\quad [D,\Theta]\omega_p
=D\omega_p,&\quad [i_v,\Theta]\omega_p=-i_v\omega_p,}
\label{extra}
\end{equation}
for any $p$-form $\omega_p$,
while $\sigma$ commutes with $D$, $\Theta$ and $i_v$.
Let us define, for $p\geq 1$,
\begin{equation}
\Omega_p=(p-1)!\sum_{n=0}^\infty\sum_{i_1\cdots i_n}(-1)^n
{\kappa_{i_1}\cdots \kappa_{i_n}\over (n+p)!}
D_{i_1}\cdots D_{i_n}\omega_p.
\label{q1}
\end{equation}
Using $D\omega_p=0$, it is immediate
to prove that $\Omega_p$ is also $D$-closed: $D\Omega_p=0$.
Moreover, notice that, if $\sigma\omega_p=0$, then $\sigma\Omega_p=0$.
The solution to our problem is ($p\geq 1$)
\begin{equation}
\delta_p=i_v \Omega_p.
\label{q2}
\end{equation}
Indeed, using the first of (\ref{extra}), one finds
\begin{equation}
D\delta_p=Di_v\Omega_p=-i_vD\Omega_p+(p+\Theta)\Omega_p=
(p+\Theta)\Omega_p=\omega_p.
\end{equation}

The integrability property that we have shown
permits to find $k$-forms $\delta_k$, $k=0,\ldots m$, such that
\begin{equation}
\omega_{k\, div}=(-1)^kD\delta_{k-1}-\sigma \delta_{k}.
\label{descentsol}
\end{equation}
One starts from $\omega_m$, which satisfies $D\omega_m=0$.
Let us set $\delta_m=0$. There exists a $\delta_{m-1}$ such that
$\omega_m=(-1)^mD\delta_{m-1}$. Then,
(\ref{descent}) gives, for $\omega_{m-1}$,
$D(\omega_{m-1}+\sigma\delta_{m-1})=0$, so that there exists a
$\delta_{m-2}$
such that $\omega_{m-1}=-\sigma\delta_{m-1}+
{(-1)}^m D\delta_{m-2}$.
Repeating this argument, one proves (\ref{descentsol}).
At the last step
one has $D(\omega_{1\, div}+\sigma\delta_1)=0$, that is solved by
$\omega_{1\, div}=-\sigma\delta_1-D\delta_0$.
Moreover, since $\sigma\omega_{1\, div}=
D\omega_{0\, div}=0$ (due to $\omega_{0\, div}=0$) and
$\sigma(\omega_{1\, div}+\sigma\delta_1)=0$,
the explicit form of the solution $\delta_0$
[formul\ae\ (\ref{q1}) and (\ref{q2}) with $\omega=\omega_{1\, div}+\sigma
\delta_1$] and the fact that $\sigma$ commutes with $i_v$
show that $\sigma\delta_0=0$, precisely as in (\ref{eqb}).
Clearly, $\delta_k$ are of order $\hbar^n$ and divergent.

Let us now define $\Delta_k=<\delta_k>_J$. We see that it is
possible to use the
cascade invariances of degrees $k=1,\ldots m-1$ to make
$\omega_{k\, div}=0$ for $k=2,\ldots m$. This is easily
seen from (\ref{cascadediv}), since
the order $\hbar^n$ divergent part $\Delta_{k\, div}$
of $\Delta_k$ coincides with $\delta_k$. At the end, we have
a $\omega_{1\, div}$ that is $\sigma$-closed and has the form
\begin{equation}
\omega_{1\, div}=-D\delta_0, \hskip .3truecm \sigma\delta_0=0.
\end{equation}
So, we have solved our problem for any $\omega_k$ except for
$\omega_1$. The final step is to use the freedom (\ref{freedom})
with $T^{(n)}=-\delta_0$, which permits to cancel $\omega_{1\, div}$
(and has no effect on the order $\hbar^n$ divergent parts of
the other $\omega_k$, as can be easily verified).

We conclude that the full set of cascade equations survives the
subtraction algorithm, allowing a complete control on the gauge-fixing
sector and proving the gauge-fixing independence of the on-shell
physical amplitudes \cite{me}.

\section{A geometrical description}
\label{geometry}
\setcounter{equation}{0}

Let us define a principal fiber bundle with ${\cal M}_{gf}$ as the
base manifold. The fibers are isomorphic to the group
of canonical transformations. The ``Lie''
algebra is the algebra of antibrackets.
To be more precise, let ${\cal I}$ denote the space of fields $\Phi$
and BRS sources $K$.
Let us give the name {\sl scalar functionals}
to those functionals
$X(\Phi,K)$ of fields and BRS sources
that transform as $X^\prime=\tilde X$ under canonical
transformations. Let ${\cal Z}$ denote the set of scalar functionals.
Let $g$ be the operator that represents the action of
a canonical transformation on a scalar functional $X$:
$gX=X^\prime=\tilde X$ and let ${\cal G}$ be the set of such
$g$'s. ${\cal G}$ is obviously a group and the map
\begin{eqnarray}
R&:& {\cal G}\times {\cal Z} \rightarrow {\cal Z}\nonumber\\
R(g,X)&=&gX
\end{eqnarray}
is a representation of ${\cal G}$.

${\cal Z}$ is obviously a vector space. Equation (\ref{caninvariance})
shows that $\omega_i\in {\cal Z}\,\,\forall i\neq 1$.
We define a product $\cdot$ in ${\cal Z}$,
represented by antibrackets
\begin{eqnarray}
\cdot \phantom{y}&:& {\cal Z}\times {\cal Z}\rightarrow {\cal Z}\nonumber\\
X\cdot Y&=& (X,Y)
\end{eqnarray}
In this way, ${\cal Z}$ becomes an algebra. To make connections with
usual notions, we notice that antibrackets replace Lie brackets and
the above
concept of algebra replaces the concept of Lie algebra. In analogy
with this,
${\cal Z}$ will be called {\sl antialgebra}. Similarly,
${\cal G}$ will be called {\sl antigroup} and replaces the Lie group.
Clearly, ${\cal Z}$ and ${\cal G}$ are infinite-dimensional.
The space ${\cal I}$ corresponds to the set of ``Lie algebra''-indices $a$,
so that $X(\Phi,K)$ corresponds to the adjoint
representation $\phi^a$ of the
Lie algebra.

Let us define
the principle fiber bundle
\begin{equation}
{\cal P}=({\cal E},{\cal M}_{gf},\pi,{\cal G}).
\end{equation}
As anticipated, ${\cal M}_{gf}$ is the base manifold. ${\cal E}$ is
such that the sections of ${\cal P}$ are obtained
by letting the canonical transformations $g\in{\cal G}$ depend
on the points $\kappa$ of the base manifold ${\cal M}_{gf}$:
$g=g(\kappa)$ and $g(\kappa)\in {\cal G}\,\,\forall \kappa$.
$\pi:{\cal E}\rightarrow {\cal M}_{gf}$ is the projection onto
the base manifold:
$\pi(g(\kappa))=\kappa$.
$\pi^{-1}(\kappa)$ is isomorphic to ${\cal G}$.

When a functional $X\in{\cal Z}$ is ${\cal M}_{gf}$-dependent,
$X=X(\Phi,K,\kappa)$, then it can be described as
a section of the fiber bundle ${\cal B}$ with typical fiber ${\cal Z}$
associated with ${\cal P}$ by the representation $R$ and
corresponds to the familiar Lie algebra-valued scalar field $\phi^a(x)$.

On the bundle ${\cal P}$, a connection can be introduced:
it is $\omega_1$. Under
a canonical transformation represented by $g$ on ${\cal Z}$,
associated to the generating functional $F_g$ (roughly
speaking, $g\sim
{\rm e}^{F_g}$), we have from
(\ref{caninvariance})
$\omega_1^\prime=\tilde \omega_1-d^\prime F$.
For an infinitesimal transformation $g=1+\varepsilon h$,
 $F_g=\Phi K^\prime+\varepsilon f_h$, we have
\begin{eqnarray}
\omega_1^\prime&=&\omega_1-\varepsilon (d f_h-(\omega_1,f_h))+\cdots
=\omega_1-\varepsilon {\cal D}f_h+\cdots\nonumber\\
X^\prime&=&gX=X-\varepsilon (f_h,X)+\cdots
\end{eqnarray}
where $X\in{\cal Z}$. The above formul\ae\ are very similar to the common
ones for a gauge-theory and justify the name ``scalar functionals'' for
the elements of ${\cal Z}$.

The effective action $\Gamma$ is a scalar functional and, moreover, it
is covariantly constant. The off-shell covariant constancy of $\Gamma$
implies the on-shell constancy, since ${\cal D}\Gamma-d\Gamma=
-{\rm ad}\, \Gamma \, \,\omega_1$, which is zero on-shell.

The above geometrical description permits to get an intuitive perception
of the fact that physical amplitudes remain independent of the gauge-fixing
parameters. Indeed, the subtraction algorithm is made of two basic
ingredients: the redefinitions of the parameters $\lambda$, that
are ``orthogonal'' to ${\cal M}_{gf}$ in the sense that they do not
depend on the point on ${\cal M}_{gf}$, and a canonical transformation.
The canonical transformation, on the other hand, is a gauge-transformation
in the principal fiber bundle ${\cal P}$, i.e.\ it is ``vertical'' with
respect to the base manifold ${\cal M}_{gf}$. For this
reason, it cannot pick up any information from
the base manifold itself. Thus, we can say that the
full subtraction algorithm is orthogonal to the manifold
${\cal M}_{gf}$ of gauge-fixing parameters.

With this, we think that we have reached a satisfactory control
on what happens to the gauge-fixing parameters.
Of course, more important is to have
control on what happens to the physical parameters
$\lambda$. An analogous covariant treatment should be introduced
as a starting point for solving the problem of
classification of predictive
nonrenormalizable quantum field theories.

Let us call ${\cal M}_{ph}$ the manifold of the physical parameters
$\{\lambda_i\}$. In general, ${\cal M}_{ph}$
is infinite dimensional (for nonrenormalizable theories), while
the dimension of ${\cal M}_{gf}$ depends on the gauge-fixing
choice. One (at least) of the parameters $\lambda$
of the classical Lagrangian (let us call it $\lambda_0$)
is peculiar, since it multiplies the gauge-invariant functional
${\cal G}_0$ that defines the propagator. So, there is no
perturbative expansion in $\lambda_0$ and the hyperplane $\lambda_0=0$
does not belong to ${\cal M}_{ph}$.

{}From now on, the differentiation (\ref{dgf}) on ${\cal M}_{gf}$
will be denoted by $d_{gf}$, while we introduce a differentiation
\begin{equation}
d_{ph}=d\lambda_i\left.{\partial\over \partial \lambda_i}\right|_{\Phi,K}
\label{dph}
\end{equation}
on the manifold ${\cal M}_{ph}$. Moreover, we write
\begin{equation}
d=d_{ph}+d_{gf},
\label{d}
\end{equation}
which is the differential operator on the manifold
${\cal M}= {\cal M}_{ph}\otimes{\cal M}_{gf}$.
Notice that, since the gauge coupling constants are not redefined
by the subtraction algorithm \cite{me}, they do not belong
to ${\cal M}$.

The redefinitions $\lambda_i\rightarrow
\lambda_i-\delta_n\lambda_i$
of the parameters $\lambda_i$ can  be described as a diffeomorphism
in ${\cal M}_{ph}$.
Indeed,
$v_n=\delta_n\lambda_i{\partial\over \partial \lambda_i}$ is a vector field
on ${\cal M}_{ph}$, $v_n\in {\cal T}{\cal M}_{ph}$ (${\cal T}$
denoting the tangent bundle)
and we can write ${\cal G}^{(n)}=
\sum_i\delta_n\lambda_i{\cal S}_i=v_n {\cal S}$ or equivalently
\begin{equation}
{\cal G}^{(n)}=l_n {\cal S},
\label{5.7}
\end{equation}
where $l_n=d i_{v_n}+i_{v_n}d$ is the Lie derivative along the vector
field $v_n$. Thus, the subtraction algorithm can be described as a
composition of diffeomorphisms $l_n$ on ${\cal M}_{ph}$,
independent of ${\cal M}_{gf}$, and canonical transformations.
(\ref{5.7}) means that
the action ${\cal S}$ transforms as a scalar under $l_n$.

We can now formulate the requirement that finitely many parameters
are sufficient to remove the divergences in this geometrical framework.
A theory is predictive if
there exists a finite dimensional
submanifold ${\cal V}\subset{\cal M}_{ph}$,
such that $v_n|_{\cal V}$ are vector fields of ${\cal V}$ $\forall n$:
\begin{equation}
\matrix{
v_n|_{\cal V}\in {\cal T}{\cal V}\, \, \forall n,&\quad {\rm dim}\,
{\cal V}<\infty.}
\end{equation}

Applying the subtraction algorithm to
a quantum field theory, one gets a set of vector fields $v_n$ on
${\cal M}_{ph}$. In particular, to each point $\lambda$ of
${\cal M}_{ph}$ a set of vectors
$\{v_1(\lambda),\ldots, v_n(\lambda), \ldots\}$ is associated. Let us
call $V(\lambda)$ the vector space spanned by these vectors.
Let us call ${\cal V}_k$ the subset of ${\cal M}_{ph}$
(it is not guaranteed that it is a manifold, but let us
suppose that it is)
where ${\rm dim}\, V(\lambda)=k$,
\begin{equation}
{\cal V}_k=\{\lambda\in {\cal M}_{ph}: \,{\rm dim}V(\lambda)=k\}.
\end{equation}
A necessary condition for
predictivity is that $\exists k<\infty$ such that ${\cal V}_k\neq
0\!\!\!\slash$.
Then, if $v_n(\lambda)\in {\cal T}{\cal V}_k$ $\forall n$ and
$\forall \lambda\in {\cal V}_k$, we can take
${\cal V}={\cal V}_k$. In general, however, this does not hold.
So, one can define a sequence of subspaces
\begin{equation}
{\cal V}_k^{(i)}=\{\lambda\in {\cal V}_k^{(i-1)}:
\, v_n(\lambda)\in {\cal T}{\cal V}_k^{(i-1)}\},
\end{equation}
where ${\cal V}_k^{(0)}={\cal V}_k$. The search stops at an $i$
such that ${\cal V}_k^{(i+1)}={\cal V}_k^{(i)}$.

Supposing that a suitable submanifold ${\cal V}$ has been found,
one can apply
the same construction to ${\cal V}$ itself, in order to
see if it is possible to {\sl reduce} the (now finite) number of
independent parameters that are necessary for predictivity.
One has to check if there exists no submanifold of ${\cal V}$
to which one can consistently restrict. A quantum field theory
with a ${\cal V}$ such that ${\rm dim}\,{\cal V}={\rm minimum}$
can reasonably be called {\sl irreducible}.
In the cases where no gauge symmetry is present,
the irreducible theory is free
(reduction to the single parameter
$\lambda_0$ plus the eventual mass), obtained by
removing all interactions. When there is some gauge-invariance, on the
other hand,
the free theory is obtained by letting the gauge coupling constant $g$
going to zero, which, however, does not correspond to
a reduction in the above sense,
since the gauge coupling constant does not belong
to the set ${\cal M}_{ph}$. So, the above definition of
irreducibility is nontrivial for gauge-theories.
The irreducible Yang-Mills theories
are the ordinary renormalizable versions. The irreducible
theory of gravity is unknown.

In the next section, we discuss some predictive (but
unphysical) toy models, that, for simplicity,
have no gauge-symmetry.

\section{Toy models}
\label{toy}
\setcounter{equation}{0}

The removal of
${\cal G}^{(n)}$ with redefinitions of $\lambda$ requires,
in general,
the presence of infinitely many $\lambda$'s, so that
predictivity is lost. The algorithm defined in \cite{me}
and generalized in the present paper is also
applyable to non-renormalizable gauge-field theories and
it is not so na\"\i ve
to pretend that in general the removal of
divergences can be performed with a finite
number of parameters (in various algorithms appeared in the literature,
instead, a preferred choice for
infinitely many parameters is hidden in some peculiar regularization
technique or in the renormalization prescription,
loosing control on the involved
arbitrariness).
Our algorithm keeps complete control on the arbitrariness introduced
in the subtraction procedure.
However, one cannot {\sl a
priori} discard the possibility that, in the context of the algorithm that we
have formulated, the
divergences of a nonrenormalizable theory
can be removed with only a {\sl finite} number of
parameters $\lambda$, while keeping a complete control
on the involved arbitrariness. The distinction between two subsets
of divergences (${\cal G}^{(n)}$ and $\sigma R^{(n)}$) that have
different roles and properties
clarifies that the problem of predictivity
only concerns ${\cal G}^{(n)}$.
As we shall see,
one can reformulate the concept of predictivity of the previous section
by means of a condition
on ${\cal G}^{(n)}$ such that it
is sufficient to redefine a
finite number of parameters in order to remove
${\cal G}^{(n)}$ itself. The remaining divergent terms
(i.e.\ $\sigma R^{(n)}$) are not
constrained to have any particular form: indeed,
infinitely many new counterterms can appear through $R^{(n)}$,
but, whatever $R^{(n)}$ is, it can always be
removed by a canonical transformation, without effects on the physical
amplitudes. This section is devoted to the search for examples
of toy models of power counting nonrenormalizable theories
in which a finite number of parameters is sufficient to remove
the divergences. Since the piece $\sigma R^{(n)}$
has no influence on predictivity, we focus for now on
non-gauge field theories, where $\sigma R^{(n)}$ is
absent: only redefinitions of the parameters $\lambda$ are
involved and no canonical transformation at all.
In a first example we show that it is possible to construct
a power counting nonrenormalizable theory that is
polynomial and such that the counterterms
are in a finite number of types and have the
same form as the terms of the classical Lagrangian. The theory is
protected by a diagrammatics that is simplified by the fact that
the propagator is off-diagonal.
Due to this, however, it is nonphysical, since
the kinetic action is not positive definite.
Nevertheless, we think that its properties with respect to the
subtraction algorithm deserve attention.
We then develop a method for producing certain
predictive nonrenormalizable theories by means of a ``change of
variables'' that has to be performed on suitable renormalizable
theories in which some composite operator is introduced, coupled
to an external source $K$.

Before entering into details, let us briefly point out some
differences between our subtraction algorithm and other algorithms
that one can find in the literature.

For example, in ref.\
\cite{tyutin} one simply redefines the action $\Sigma$
by subtracting, order by order, the
divergent part $\Gamma^{(n)}_{div}$, i.e.\
\begin{equation}
\Sigma_n=\Sigma_{n-1}-\Gamma^{(n)}_{div}.
\label{uno}
\end{equation}
The dimensional
regularization technique is used.
The algorithm (\ref{uno}) breaks the master equation, however one can
prove that this breaking, at least in the dimensional
regularization framework, is under control. At the end,
the ``renormalized'' action $\Sigma_\infty$ satisfies
the classical master equation
$(\Sigma_\infty,\Sigma_\infty)=0.$
The algorithm (\ref{uno}), apart from the restriction on
the regularization technique, is completely general.
The ``philosophy of the method'' is the following:
whenever a divergent
term is found, it has to be
subtracted away. One does not wonder whether the divergent
term is of a new type or not.
The limit of the algorithm, however,
is that one does not have a direct control
on the arbitrariness of the procedure.

On the other hand, if one is not satisfied with the simple subtraction
of the divergent terms, but one requires this subtraction to be implemented by
a redefinition of parameters and fields,
then the classical Lagrangian is, in general, demanded to possess
an infinite number of parameters $\lambda$.
In this case, predictivity is lost.

Now, we go on by reformulating the predictivity requirement of the end
of section \ref{geometry} in more concrete terms.
Suppose that the $\lambda_i$ are certain functions
of a finite number of new parameters $\alpha_j$, i.e.\
\begin{equation}
{\cal L}_{class}={\cal L}_{class}(\phi,\alpha)=
\sum_i \lambda_i(\alpha) \,{\cal G}_i.
\label{pou}
\end{equation}
Then, $\delta_n\lambda_i$ are also functions of $\alpha$:
\begin{equation}
{\cal G}^{(n)}=\sum_i \delta_n\lambda_i (\alpha)\,{\cal S}_i(\lambda(\alpha)).
\end{equation}
If there
exist suitable $\Delta_n\alpha_j$, of order
$\hbar^n$, such that
\begin{equation}
{\cal G}^{(n)}=\sum_j
\Delta_n\alpha_j{\partial S\over \partial\alpha_j}
=\sum_{j,i}
\Delta_n\alpha_j {\partial \lambda_i(\alpha)\over \partial \alpha_j}
{\cal S}_i,
\label{nine}
\end{equation}
then it is possible to cancel ${\cal G}^{(n)}$ by
simply redefining the $\alpha_j$.
Indeed,
\begin{eqnarray}
\Sigma_{n-1}(\Phi,K,\alpha_j-\Delta_n\alpha_j)&=&
\Sigma_{n-1}(\Phi,K,\alpha)
-\sum_j\Delta_n\alpha_j{\partial\Sigma_{n-1}\over \partial\alpha_j}
+{\cal O}(\hbar^{n+1})\nonumber\\&=&
\Sigma_{n-1}(\Phi,K,\alpha)
-\sum_j\Delta_n\alpha_j{\partial S\over \partial\alpha_j}
+{\cal O}(\hbar^{n+1})\nonumber\\&=&
\Sigma_{n-1}-{\cal G}^{(n)}+{\cal O}(\hbar^{n+1}).
\end{eqnarray}
The term $\sigma R^{(n)}(\Phi,K)$ is then cancelled in the known way
by a canonical transformation.

Condition (\ref{nine}) is equivalent to
\begin{equation}
\delta_n\lambda_i(\alpha)=\sum_j\Delta_n\alpha_j
{\partial \lambda_i(\alpha)\over \partial\alpha_j}.
\label{fundamentalequations}
\end{equation}
These are nontrivial conditions (they should be satisfied
for any $n$) on the functions $\lambda_i(\alpha)$.
It is clear that the power-counting
renormalizable theories trivially satisfy
condition (\ref{fundamentalequations}): the sets
$\{\lambda_i\}$ and $\{\alpha_j\}$ coincide.

The contact with the geometrical formulation of the predictivity requirement
given at the end of the previous section is that $\alpha_j$
are coordinates on the finite dimensional manifold ${\cal V}$ and
\begin{equation}
\lambda_i=\lambda_i(\alpha_j)
\end{equation}
 are the equations embedding ${\cal V}$  in ${\cal M}_{ph}$, while
 (\ref{fundamentalequations}) corresponds to the condition
 $v_n(\lambda)\in {\cal TV}\,\,\forall\lambda\in{\cal V}$,
 equivalent to $v_n(\lambda(\alpha))\in {\cal TV}\,\,\forall\alpha$.

We emphasize again that a possible source of insight is to investigate
this kind of ``stability'' with respect to the subtraction algorithm, or a
stability ``in the sense of the correspondence principle''.
Instead, we never refer to stability with respect to the renormalization
group.

The next task is to elaborate some toy models
of nonrenormalizable theories that
satisfy (\ref{fundamentalequations}) in order to show that we are
dealing with something nonempty and nontrivial.

Let us start with a polynomial theory. It consists of two scalar
fields $\phi_1$ and $\phi_2$ with (nonpositive definite)
kinetic action $-\phi_1\Box \phi_2+m^2\phi_1\phi_2$.
Consider the Lagrangian
\begin{equation}
{\cal L}_{class}(\phi_1,\phi_2)=-\phi_1\Box \phi_2+m^2\phi_1\phi_2+
{\lambda\over 4}
\phi_1^2\phi_2^2.
\label{prima}
\end{equation}
It is an exotic $\phi^4$-type theory. We work with the dimensional
regularization technique. It is clear that the theory is
ultraviolet renormalizable (but not finite):
there exist suitable constants $Z$, $\delta m^2$ and
$Z_\lambda$ such that the renormalized Lagrangian
\begin{equation}
{\cal L}_{ren}(\phi_1,\phi_2)=-Z\phi_1\Box \phi_2+
Z(m^2+\delta m^2)\phi_1\phi_2+{\lambda Z_\lambda\over 4}
Z^2\phi_1^2\phi_2^2,
\end{equation}
gives a convergent generating functional $W$ of the connected Green
functions.
Let us now introduce a $\phi^5$-type term, precisely,
\begin{equation}
{\cal L}_{class}(\phi_1,\phi_2)=-\phi_1\Box \phi_2+
m^2\phi_1\phi_2+{\lambda\over 4}
\phi_1^2\phi_2^2+{\alpha\over 4!}\phi_2\phi_1^4.
\end{equation}
Despite the appearance, only a finite number
of types of divergent graphs are generated,
because the diagrammatics is very simple.
One can easily check that there are two-loop divergent diagrams
with three external $\phi_1$-legs. The dimensions are such that
the power in the external momenta is two.
The required counterterms have the form $\phi_1^2\Box\phi_1$
and $\phi_1^3$.
Moreover, one loop divergent diagrams with six external
$\phi_1$-legs can easily been constructed, so that a counterterm
$\phi_1^6$ is also required.
For a reason that will be clear in a moment, let us also introduce
a $\phi_2^3$ vertex.
One is thus lead to consider the Lagrangian
\begin{equation}
{\cal L}_{class}(\phi_1,\phi_2)=-\phi_1\Box \phi_2+m^2\phi_1\phi_2+
{\eta \over 3!}\phi_1^3+{\lambda\over 4}
\phi_1^2\phi_2^2+{\alpha\over 4!}\phi_2\phi_1^4+{\beta\over 3!}\phi_1^2
\Box\phi_1+{\gamma\over 6!}\phi_1^6+{\zeta\over 3!}\phi_2^3,
\label{ex1}
\end{equation}
where $\eta$, $\lambda$, $\alpha$, $\beta$, $\gamma$ and $\zeta$ are
independent
(and ``small'') coupling constants.
The diagrammatics is so simple that it is easy to check
that no other counterterms are generated. Indeed, the
kinetic action $-\phi_1\Box \phi_2+m^2\phi_1\phi_2$ has been
chosen precisely to simplify the diagrammatics.
Let us give the explicit proof. Let $G$ denote a graph
with $E_1$ external $\phi_1$-legs, $E_2$ external $\phi_2$-legs,
$I$ internal legs and $L$ loops.
Let $n_1$, $n_2$, $n_3$, $n_4$, $n_5$  and $n_6$ denote the number of vertices
of the forms $\phi_1^3$,
$\phi_1^2\phi_2^2$, $\phi_1^4\phi_2$, $\phi_1^2\Box\phi_1$,
$\phi_1^6$ and $\phi_2^3$, respectively. Finally, let $\omega(G)$ denote
the superficial degree of divergence of the graph $G$. We have
\begin{eqnarray}
I+E_1&=&3n_1+2n_2+4n_3+3n_4+6n_5,\nonumber\\
I+E_2&=&2n_2+n_3+3n_6,\nonumber\\
L&=&I-n_1-n_2-n_3-n_4-n_5-n_6+1.
\label{bia}
\end{eqnarray}
The first two formul\ae\ give the total numbers of $\phi_1$-
and $\phi_2$-legs. Each propagator connects a $\phi_1$-leg with
a $\phi_2$-leg.
The superficial degree of divergence turns out to be
\begin{eqnarray}
\omega(G)&=&4L-2I+2n_4=2I-4n_1-4n_2-4n_3-2n_4-4n_5-4n_6+4\nonumber\\&=&
4+2I-{2\over 3}(I+E_1+2I+2E_2)-2n_1
=4-{2\over 3}(E_1+2E_2)-2n_1.
\end{eqnarray}
As we see, $\omega(G)$ is bounded. We have to show that no new
counterterm is required.

For $\omega(G)=0$ there are two possibilities:

i) $n_1$=0. In this case, we have $E_1=6$, $E_2=0$,
or $E_1=4$, $E_2=1$, or $E_1=2$, $E_2=2$, or $E_1=0$, $E_2=3$;

b) $n_1=1$. Now, it can only be $E_1=3$, $E_2=0$, or $E_1=1$, $E_2=1$.

Instead, $\omega(G)=2$
is only consistent with $E_1=3$, $E_2=0$, or with $E_1=1$, $E_2=1$
at $n_1=0$.
All these divergences have the form of the quadratic part of the
Lagrangian or
of the vertices.
Equations (\ref{fundamentalequations}) are trivially
satisfied [in the present case, the $\alpha$-parameters of (\ref{pou})
coincide with the $\lambda$-parameters]. Thus the theory
is predictive, nonrenormalizable, nonfinite, polynomial and has a nonpositive
definite kinetic action.

It we set $\zeta=0$, the theory remains predictive with five parameters
($\eta$, $\lambda$, $\alpha$, $\beta$ and $\gamma$).
Indeed, when $\phi_2^3$ is absent ($n_6=0$),
eq.\ (\ref{bia}) gives $E_1\geq E_2$, so that the solution $E_1=0$,
$E_2=3$ has to be discarded: the $\phi_2^3$ vertex is not radiatively
generated, if it is initially absent.

We can do even more, namely we can fix some parameters as suitable functions
of the others, while preserving predictivity. In particular, for
a reason that we shall discuss in a moment,
the theory with Lagrangian
\begin{eqnarray}
{\cal L}_{class}(\phi_1,\phi_2,\lambda,\alpha,m^2)&=&
-\phi_1\Box \phi_2+m^2\phi_1\phi_2-\alpha m^2 \phi_1^3+
{\lambda\over 4}\phi_1^2\phi_2^2\nonumber\\&&-{\lambda\alpha\over 2}
\phi_2\phi_1^4+\alpha\phi_1^2
\Box\phi_1+{\lambda \alpha^2\over 4}\phi_1^6
\label{1the}
\end{eqnarray}
is predictive. Now eqs. (\ref{fundamentalequations}) are still
verified, but in a nontrivial way. The independent parameters
have been reduced to three: $m$, $\lambda$ and $\alpha$.
To check that the theory (\ref{1the}) is indeed predictive, let us introduce
the sources explicitly:
\begin{eqnarray}
{\cal L}_{class}(\phi_1,\phi_2)&=&-\phi_1\Box \phi_2+
m^2\phi_1\phi_2-\alpha m^2 \phi_1^3+{\lambda\over 4}
\phi_1^2\phi_2^2-{\lambda\alpha\over 2}\phi_2\phi_1^4
\nonumber\\&+&\alpha\phi_1^2
\Box\phi_1+{\lambda \alpha^2\over 4}\phi_1^6+J_1\phi_1+J_2\phi_2.
\label{2the}
\end{eqnarray}
We now perform the following change of variables in the functional integral
(the Jacobian determinant being one)
\begin{equation}
\matrix{\varphi_1=\phi_1,\cr
\varphi_2=\phi_2-\alpha \phi_1^2,}
\end{equation}
so that (\ref{2the}) can be rewritten as
\begin{equation}
{\cal L}_{class}^\prime=
-\varphi_1\Box \varphi_2+m^2\varphi_1\varphi_2
+{\lambda\over 4}\varphi_1^2\varphi_2^2
+J_1\varphi_1+J_2(\varphi_2+\alpha\varphi_1^2).
\label{3the}
\end{equation}
Now the Lagrangian has a quite honest aspect, however, the sources do
not appear in the conventional way $J_1\varphi_1+J_2\varphi_2$.
Nevertheless, let us call $W[J_1,J_2,\lambda,\alpha,m^2]$
the generating functional of the connected
Green functions for the theory (\ref{2the}) or (\ref{3the}).
We can write
\begin{equation}
W[J_1,J_2,\lambda,\alpha,m^2]=
\tilde W[J_1,J_2,K,\lambda,m^2]|_{K=\alpha J_2},
\end{equation}
where $\tilde W[J_1,J_2,K,\lambda,m^2]$
is the generating functional of the theory with Lagrangian
\begin{equation}
\tilde{\cal L}_{class}=
-\varphi_1\Box \varphi_2+m^2\varphi_1\varphi_2+
{\lambda\over 4}\varphi_1^2\varphi_2^2
+J_1\varphi_1+J_2\varphi_2+K\varphi_1^2.
\label{poi}
\end{equation}
This is simply the renormalizable theory (\ref{prima}) with the introduction
of the composite operator $\varphi_1^2$, with source $K$.
If $G$ is a graph with $E_1$ external $\varphi_1$-legs,
$E_2$ external $\varphi_2$-legs, $I$ internal legs,
$L$ loops and $E_K$ external $K$-legs, the superficial degree of
divergence $\omega(G)$ turns out to be
$\omega(G)=4-2E_1$. Moreover, it is easy to show that
$E_1=E_2+2E_K$. Thus, the only divergent graphs
[$\omega(G)=0$ and $\omega(G)=2$] have the form of the terms
of $\tilde {\cal L}_{class}$ (\ref{poi}). Consequently,
there exist constants $Z$, $\delta m^2$, $Z_\lambda$ and $Z_\alpha$ such
that the renormalized Lagrangian
\begin{equation}
\tilde {\cal L}_{ren}=
-Z\varphi_1\Box \varphi_2+Z(m^2+\delta m^2)
\varphi_1\varphi_2+{\lambda Z_\lambda\over 4}Z^2\varphi_1^2\varphi_2^2
+J_1\varphi_1+J_2\varphi_2+Z_\alpha Z^{1/2} K\varphi_1^2,
\end{equation}
gives a finite generating functional $\tilde W[J_1,J_2,K,\lambda,m^2]$.
Setting $K=\alpha J_2$ and going back with the
renormalized inverse
change of variables, namely
\begin{equation}
\matrix{\phi_1=\varphi_1,\cr
\phi_2=\varphi_2+\alpha Z_\alpha Z^{1/2}\varphi_1^2,}
\end{equation}
the theory
\begin{eqnarray}
{\cal L}_{ren}(\phi_1,\phi_2,\lambda,\alpha,m^2)&=&
-Z\phi_1\Box \phi_2+Z(m^2+\delta m^2)\phi_1\phi_2
-\alpha Z_\alpha Z^{3/2} (m^2+\delta m^2)\phi_1^3\nonumber\\&&+
{\lambda Z_\lambda \over 4}
Z^2\phi_1^2\phi_2^2-{\lambda\alpha Z_\lambda Z_\alpha\over 2}
Z^{5/2}\phi_2\phi_1^4+\alpha Z_\alpha Z^{3/2}\phi_1^2
\Box\phi_1\nonumber\\&&
+{\lambda \alpha^2 Z_\lambda Z_\alpha^2\over 4}Z^3\phi_1^6
+J_1\phi_1+J_2\phi_2,
\label{4the}
\end{eqnarray}
corresponds to a finite generating functional
$W_{ren}[J_1,J_2,\lambda,\alpha,m^2]$.
Notice that (\ref{2the}) and (\ref{3the}) are the same theory, while
(\ref{3the}) and (\ref{poi}) are different theories.
We see that ${\cal L}_{ren}(\phi_1,\phi_2,\lambda,\alpha,m^2)=
{\cal L}_{class}(Z^{1/2}\phi_1,Z^{1/2}\phi_2,
\lambda Z_\lambda,\alpha Z_\alpha,m^2+\delta m^2)$.

Let us rewrite ${\cal L}_{class}$ and ${\cal L}_{ren}$ in a form that is more
similar to (\ref{l1}). Let us introduce a parameter $\zeta$ in front
of the kinetic Lagrangian. $\zeta$ is not ``small'', i.e.\ we do
not make a perturbative expansion in $\zeta$. We write
\begin{eqnarray}
{\cal L}_{class}(\zeta,\lambda,\alpha)&=&-\zeta\phi_1\Box \phi_2+
\zeta m^2\phi_1\phi_2
-\zeta \alpha m^2 \phi_1^3
+{\lambda\over 4}
\phi_1^2\phi_2^2
\nonumber\\&-&{\lambda\alpha\over 2}\phi_2\phi_1^4+\zeta\alpha\phi_1^2
\Box\phi_1+{\lambda \alpha^2\over 4}\phi_1^6.
\end{eqnarray}
Then, there exist factors $\tilde Z_\zeta$, $\tilde\delta m^2$,
$\tilde Z_\lambda$ and
$\tilde Z_\alpha$
such that the renormalized Lagrangian is
\begin{equation}
{\cal L}_{ren}(\zeta,\lambda,\alpha,m^2)={\cal L}_{class}
(\zeta \tilde Z_\zeta, \lambda \tilde Z_\lambda,\alpha \tilde
Z_\alpha,m^2+\tilde\delta m^2).
\end{equation}
In other words, ${\cal L}_{ren}$ is obtained from ${\cal L}_{class}$
precisely with suitable redefinitions of the four parameters
$\zeta$, $m^2$, $\lambda$ and $\alpha$. Restoring $\zeta=1$, one
has $\tilde Z_\zeta|_{\zeta=1}=Z$,
$\tilde\delta m^2|_{\zeta=1}=\delta m^2$,
$\tilde Z_\lambda|_{\zeta=1}=Z_\lambda Z^2$
and $\tilde Z_\alpha|_{\zeta=1}=Z_\alpha Z^{1/2}$.

What we have elaborated is a method for constructing certain
nonrenormalizable predictive theories from renormalizable ones.
It is worth stopping for a moment and giving a clear
description of this method.
One starts from a renormalizable theory
of certain fields $\Phi$. Let the corresponding sources be
denoted by $J$: in the functional integral, $J$ only
appear in the linear term $J\Phi$ that is added to the action.
Then, one introduces some suitable composite operators ${\cal O}(\Phi)$,
coupled to external sources $K$.
So, $J$ and $K$ appear in the form $J\Phi+
K{\cal O}(\Phi)$.
For simplicity, let us adopt the convention
that the sum of the Lagrangian plus
$J\Phi+K{\cal O}(\Phi)$ is still called the ``Lagrangian''.
Things have to be arranged in
such a way that only counterterms that are linear in $K$
are generated, i.e.\ such that the
renormalized Lagrangian is still linear in $K$.
Afterwards, one identifies $K$
with $J$: $K=\alpha J$, $\alpha$ being a parameter that, in general,
is negatively dimensioned.
The sources $J$ appear now
in the form $J[\Phi+\alpha {\cal O}(\Phi)]$, that can be
turned to the standard form
$J\tilde\Phi$ by a change of variables $\tilde\Phi=\Phi+
\alpha {\cal O}(\Phi)$.
A simple diagrammatic analysis \cite{veltman}
shows that if the functional integral is convergent
in the initial variables, then it is also convergent in the new
variables.
In general, the
new Lagrangian $\tilde {\cal L} (\tilde \Phi)$
contains nonrenormalizable vertices, due to the negative dimension of $\alpha$.
$\tilde {\cal L}(\tilde\Phi)$ describes the physical
content of the new theory.
In particular, the new field $\tilde\Phi$ is the elementary field of the
theory $\tilde{\cal L}(\tilde\Phi)$ and a composite field of the
theory ${\cal L}(\Phi)$. Viceversa for $\Phi$.
Since the generating functional $W[J,K]$ of the initial theory
was convergent [even in presence of the composite operators ${\cal O}(\Phi)$],
an identification between $J$ and $K$ produces a new convergent
generating functional $\tilde W[J]=W[J,\alpha J]$,
which is, in fact, the generating functional of the new theory
$\tilde {\cal L}(\tilde \Phi)$.
The new theory is predictive, as the initial
one. Moreover, if the initial theory is not finite,
then the new theory is also not finite.

Notice that it is more convenient to deal with $W$
than with the effective action $\Gamma$ (i.e.\
the Legendre transform
of $W$ with respect to $J$),
since the Legendre transform changes nontrivially
when $K$ is identified with $J$
(and, in fact, the physical meaning of the effective action changes
correspondingly). The new effective action $\tilde\Gamma$ is convergent,
since it is the Legendre transform of a convergent functional $\tilde W[J]$.
$\tilde\Gamma$ is the generating functional of the irreducible graphs
of the theory described by $\tilde {\cal L}(\tilde\Phi)$.

Now, the reason why we chose a kinetic action of the form
$-\phi_1\Box \phi_2+m^2\phi_1\phi_2$ is clear:
it was to avoid counterterms
quadratic in $K$, that would in general be required
when introducing a $\phi^2$-type composite operator
coupled to the source $K$.
Only if linearity in $K$ is preserved, we can safely
apply the above procedure by identifying $K$ with $J$.
Instead, if there are counterterms that are nonlinear in $K$, let us say
quadratic, then the identification $K=\alpha J$ produces quadratic terms
in $J$. Then, it is easy to check that the convergence of $W[J,\alpha J]$
only means that the connected diagrams converge, while the connected
{\sl irreducible} ones do not converge, in general. Indeed,
due to the nonlinearity of ${\cal L}$ in $J$, the Legendre transform
$\Gamma$ of $W$ is {\sl not} the set of connected irreducible graphs.

To further illustrate the method,
let us see what happens when
setting $K=\alpha J_1$ in $\tilde W[
J_1,J_2,K,\lambda,m^2]$.
This means that we are considering a theory described by the
Lagrangian
\begin{equation}
{\cal L}_{class}^\prime(\varphi_1,\varphi_2)=-\varphi_1\Box \varphi_2+
m^2\varphi_1\varphi_2+
{\lambda\over 4}
\varphi_1^2\varphi_2^2+
J_1(\varphi_1+\alpha\varphi_1^2)+J_2\varphi_2.
\end{equation}
Then, we perform the change of variables
\begin{equation}
\matrix{\phi_1=\varphi_1+\alpha\varphi_1^2,\cr
\phi_2=\varphi_2.}
\end{equation}
We assume to use the dimensional regularization technique, so that
the Jacobian determinant is still trivial.
Let $\varphi_1(\phi_1,\alpha)$ be the inverse of
$\phi_1=\varphi_1+\alpha\varphi_1^2$
(to be intended as a power series in $\alpha$).
We get a theory described by
\begin{equation}
{\cal L}_{class}(\phi_1,\phi_2,\lambda,\alpha)=
-\varphi_1(\phi_1,\alpha)\Box\phi_2+m^2
\varphi_1 (\phi_1,\alpha)\phi_2+
{\lambda\over 4}\varphi_1^2(\phi_1,\alpha)\phi_2^2+
J_1\phi_1+J_2\phi_2.
\end{equation}
In this example the Lagrangian is nonpolynomial. The first
nonrenormalizable vertices are
\begin{equation}
-5m^2\alpha^3\phi_1^4\phi_2+
\alpha \phi_1^2\Box\phi_2-{1\over 2}\lambda\alpha\phi_1^3\phi_2^2.
\end{equation}
Nevertheless, the coefficients of the infinitely many nonrenormalizable
counterterms are related in such a way that eqs.\
(\ref{fundamentalequations})
are satisfied. Indeed, reasoning in a similar way as before,
setting $\phi_1=\varphi_1+\alpha Z_\alpha Z^{1/2}\varphi_1^2$ and
$\phi_2=\varphi_2$, we find that the
``renormalized'' Lagrangian is
\begin{eqnarray}
{\cal L}_{ren}(\phi_1,\phi_2)&=&-Z\varphi_1(\phi_1,\alpha Z_\alpha
Z^{1/2})
\Box \phi_2+Z(m^2+\delta m^2)
\varphi_1 (\phi_1,\alpha Z_\alpha Z^{1/2})\phi_2
\nonumber\\&+&{\lambda Z_\lambda \over 4}
Z^2\varphi_1^2(\phi,\alpha Z_\alpha Z^{1/2})\phi_2^2
\nonumber\\&=&
{\cal L}^\prime_{class}(Z^{1/2}\phi_1,Z^{1/2}\phi_2,\lambda Z_\lambda,
\alpha Z_\alpha,m^2+\delta m^2).
\end{eqnarray}

Up to know, we do not know whether
the nonpositive definiteness of the kinetic action
is an essential requirement for the above mechanism to work.
Surely, this aspect deserves attention.

Let us conclude with some brief remarks and comments about the
concept of predictivity that is formulated in this and in the previous
section of the paper.

One can wonder if there is some hidden symmetry that protects the above
theories and makes it possible to have predictivity in presence of
power-counting nonrenormalizable interactions.
Renormalizable
theories are also protected by a ``symmetry'', that is
power counting. On the other hand, the ``symmetry'' that
protects some of the predictive nonrenormalizable
theories that we exhibited is purely
``diagrammatical'', i.e.\
the impossibility of constructing many divergent graphs.
This was the criterion with which we constructed the theory
(\ref{ex1}): in particular, the quadratic part $-\phi_1\Box\phi_2+m^2
\phi_1\phi_2$
was responsible of the limited number of divergent graphs. As one
can see, there is no need of a sophisticated symmetry (like a local symmetry
or a supersymmetry) to have predictivity. This fact suggests that
the set of predictive theories is not so small. In the general case,
we do not possess, up to now, any description of the
``symmetry principle'' contained in
equations (\ref{fundamentalequations}),
simpler than eq.s (\ref{fundamentalequations}) themselves.

What about the finite counterterms that one can attach to the
divergent ones? Eq.\ (\ref{pou}) imposes
a relation among the coefficients $\lambda_i$ of the terms of the
classical Lagrangian ${\cal L}_{class}(\phi)$
and eqs.\ (\ref{fundamentalequations})
express the consistency between $\lambda_i$ and
the coefficients $\delta_n\lambda_i$
of ${\cal G}^{(n)}(\phi)
=\sum_i\delta_n\lambda_i{\cal G}_i(\phi)$.
Now, it is not permitted to add finite terms
$\sum_i f_i{\cal G}_i(\phi)$, with arbitrary finite coefficients
$f_i$, since such coefficients are in general
infinitely many and thus predictivity is lost. Stated
differently,
such an addition of $\sum_i f_i{\cal G}_i(\phi)$
is equivalent to redefine $\delta_n\lambda_i$
as $\delta_n\lambda_i-f_i$ and consequently relations
(\ref{fundamentalequations})
are in general not preserved: the $f_i$ cannot be
reabsorbed as a redefinition of $\alpha_j$ and at the subsequent
orders the divergent terms are out of control.
Thus, the finite terms that we are allowed to
add are not completely
arbitrary, rather they are restricted to be of the form
\begin{equation}
\sum_{j,i}f_j{\partial\lambda_i(\alpha)\over \partial\alpha_j}
{\cal G}_i(\phi),
\label{restriction}
\end{equation}
which corresponds to the shifts $\Delta_n\alpha_j\rightarrow
\Delta_n\alpha_j-f_j$.
Moreover, the $f_i$ should be independent of the gauge-fixing parameters,
otherwise an accidental dependence on these parameters would
be introduced.

In order to compare this situation with a more familiar one,
let us go back to quantum gravity. Goroff and Sagnotti proved
\cite{sagnotti} that Einstein gravity is not two-loop finite, because of a
divergent term equal to
\begin{equation}
{1\over \varepsilon}{209\over 2880 (4\pi)^4}\sqrt{g}{R_{\mu\nu}}^{\rho\sigma}
{R_{\rho\sigma}}^{\alpha\beta}{R_{\alpha\beta}}^{\mu\nu},
\label{2loopgr}
\end{equation}
where $\varepsilon=4-d$.
This divergence is not reabsorbable with a redefinition of the
metric tensor and so, one is forced to introduce
a higher derivative term in the Lagrangian.
However, let us suppose, for a moment, that the divergence (\ref{2loopgr})
was miraculously absent. In other words, let us consider a theory in
which some divergent term ${\cal D}$ is {\sl a priori}
allowed, but effectively absent. Moreover, let us suppose that the absence
of ${\cal D}$ is fundamental for finiteness.
Although ${\cal D}$ is not
{\sl a priori} discarded, one should avoid
to introduce it as a finite gauge-invariant
counterterm, since finiteness would be destroyed.
This means that one has to provide an {\sl ad hoc} restriction on the
finite counterterms in order to avoid terms like ${\cal D}$.
Indeed, the addition of such finite terms
corresponds to an unwanted modification
of the initial Lagrangian, in particular to the introduction
of new coupling constants and the idea of finiteness would be meaningless.

More or less the same situation characterizes the predictive
theories that we have defined. A restriction on the finite terms,
like (\ref{restriction}),
is necessary, otherwise predictivity is lost.
Strictly speaking,
an analogous restriction is also present in the case of renormalizable
theories: one restricts the admissible finite counterterms with
the criterion of power-counting. In other words, the finite
counterterms have to respect the ``symmetry'' that protects the theory,
otherwise an infinite degree of arbitrariness is introduced.
Since the ``symmetry'' that protects our predictive theories
is represented by equations (\ref{fundamentalequations}),
it is quite natural to restrict the finite counterterms correspondingly.

Sometimes, when dealing with predictivity, one
pays attention to the ``number of
measurements'' that is necessary to fix the theory.
Instead, we have never mentioned this, so far.
Rather, our criterion for predictivity is the correspondence
principle. Indeed, the number of measurements that fix uniquely the
theory can be a misleading concept. Since the physical amplitudes
of our predictive theories are supposed to
depend on a finite number of parameters $\lambda$
plus the gauge-couplings \cite{me},
it is clear that a finite number of measurements is sufficient
to fix them and consequently determine the theory uniquely.
However, the classical Lagrangian can be nonpolynomial, i.e.\
it can contain an infinite number of terms,
so that one could say that it is necessary to check experimentally
the consistency of an
infinite number of interactions with the coefficients
$\lambda_i(\alpha)$ of the corresponding vertices in
the Lagrangian.
Thus, at least when
the theory is nonpolynomial, an infinite number of measurements
would be necessary.
However, again strictly speaking, one should conclude that an
infinite number of measurements is also necessary when the theory is
renormalizable. Indeed, in that case,
there are infinitely many possible
Lagrangian terms (the power-counting
nonrenormalizable ones) that are multiplied ``by the coefficient $0$'' and
testing these values $0$ experimentally
would be a check of the power counting criterion, but it would require
infinitely many measurements.
Instead, the usual power-counting criterion is an {\sl assumption}
and our concept of predictivity can be viewed as a generalization of it.
If one accepts
hat the good criterion is the
correspondence principle, then one has a {\sl conceptual} restriction
on the set of physical theories; {\sl after} that, the experimental
measurements further project this set onto the set of
realized theories.

Some of the examples that we have constructed are suitable
``change of variables'' of renormalizable theories, some others,
like (\ref{ex1}), are not. What is the general way of proceeding,
at least in principle, to investigate whether or not a given
classical nonrenormalizable theory can be made predictive?
First, we have to specify what we mean by ``classical theory'' in
this context. Indeed, the functions $\lambda_i(\alpha)$ are
not known {\sl a priori}, so that the classical Lagrangian
is itself not known. Thus the classical theory is identified by
the field content and the gauge symmetry. The classical
Lagrangian should be constructed {\sl together} with
the quantum theory. It is identified as the classical Lagrangian
that permits the implementation of the correspondence principle
(with a finite number of free parameters $\lambda$).
At the present stage, we cannot say more about the set
of solutions: it can be empty or there can be a single solution
or eventually more solutions. The problem of
classifying the quantum field theories that are predictive according
to (\ref{fundamentalequations}) is surely deserving of interest.
One should start from the most general classical Lagrangian
${\cal L}_{class}=\sum_i\lambda_i{\cal G}_i(\phi)$.
Then one should compute the one
loop divergences ${\cal G}^{(1)}(\phi)=
\sum_i\delta_1\lambda_i{\cal G}_i(\phi)$. Making a suitable ansatz
about the number of parameters $\alpha_j$, one should then solve
eqs.\ (\ref{fundamentalequations}) for $\lambda_i(\alpha)$.
If the solution is a deep property of the theory, eqs.\
(\ref{fundamentalequations}) should be solved for any $n$.

\section{Conclusions}
\label{conclusions}
\setcounter{equation}{0}

In this paper we have extended a previously formulated subtraction algorithm
and we have reached a satisfactory control on the effects of the
subtraction procedure on the ``easy part'' of the problem, namely
the gauge-fixing sector. This investigation can be useful, to our opinion,
for establishing a convenient framework for the study of
the ``difficult part'' of the problem, to get a
satisfactory knowledge on the
effects of the subtraction algorithm on the physical parameters.
We should classify the theories that can be
quantized with a {\sl finite} number of parameters, a problem that
surprisingly has not been considered with sufficient attention, so far.
We have noticed that, in order to do this,
one has to determine the classical Lagrangian
and the full quantum theory {\sl contemporarily}. The classical
Lagrangian is a sum of (possibly) infinitely many terms, whose
coefficients are suitable functions of a finite number of parameters.
Lagrangians of this type could be furnished as effective Lagrangians of
some more fundamental theory (like string theory).
The tree level part of the
effective Lagrangian could be considered as
the classical Lagrangian of a
quantum field theory. Then, one should check
whether conditions (\ref{fundamentalequations})
are satisfied or not.

\end{document}